\documentclass[aps,preprint,12pt,tightenlines]{revtex4}%
\usepackage{amsfonts}
\usepackage{amsmath}
\usepackage{amssymb}
\usepackage{graphicx}
\usepackage{color}%
\providecommand{\U}[1]{\protect\rule{.1in}{.1in}}

\begin{document}
\preprint{ }
\title[ ]{A Coherent and Unified Single Particle Description of the Integer and Fractional Quantum Hall Effects}
\author{C. S. Unnikrishnan}
\affiliation{Tata Institute of Fundamental Research, Homi Bhabha Road, Mumbai 400005, India}

\begin{abstract}
There are compelling reasons to seek a new coherent description of the Quantum
Hall Effects (QHE). The theoretical descriptions of the `Integer' (IQHE) and
the `Fractional' (FQHE) quantum Hall effects are very different at present,
despite their remarkable phenomenological similarity. In particular, the
fractional effect invokes multi-particle dynamics and collective phenomena in the presence of a dominant
Coulomb interaction, in a complex hierarchical scheme, whereas the integer
effect is mostly a simple weakly interacting single particle scenario. The
experimental situation, in contrast, shows that both the effects
appear seamlessly, intermingling progressively, as either the magnetic field
or the carrier density is monotonically varied. I argue and prove that a
crucial physics input that is missing in the current theories is the relativistic 
gravitational contribution of all the matter-energy in the Universe. The dynamically induced
relativistic gravitational potentials play a startling role to modify the quantum degeneracy, by coupling to the mass of electrons in the noninertial cyclotron orbitals.
The key point is that the quantum degeneracy of Landau levels, due to the applied magnetic field,
is modified by the relativistically induced cosmic gravitomagnetic field, thereby
making the degeneracy dependent on the number density of the
electrons. I successfully derive the main characteristics and the full
sequence of both IQHE and FHQE in a seamless unified single particle scenario,
without any quasiparticles, particle-flux composites, or extraneous postulates. Apart from correctly
reproducing all the observed filling factors in the IQHE and the FQHE for the
filling factors $\nu\geq1/3$, this new theory of the QHE based on the cosmic
gravitational effects has the natural explanation for the absence of the QHE
at even fractions for $\nu<1$. Further, there is a consistent description of the
edge state physics, for both charge transport and thermal transport, in the
FQHE states. The gravitational paradigm shows clearly the physical reason for
the phenomenological success of the effective theories with the quasiparticles
like the Composite Fermions.

\end{abstract}
\startpage{1}
\endpage{102}
\maketitle
\tableofcontents

\begin{quote}
\newpage

\texttt{The true theory of the FQHE is yet to come, hopefully within the next
20 years. (M. I. Dyakonov, 2002, in arXiv:cond-mat/0209206)}
\end{quote}

\section{Preface}

The phenomenological equivalence of the integer and fractional quantum Hall effects \cite{QHE-book}, in a wide range of particle densities and magnitudes of the applied magnetic field, suggests the equivalence of the underlying physics and invites a unified microscopic theoretical description. In this paper, I discuss a coherent and unified single particle quantum theory
of the integer and fractional quantum Hall effects, without quasiparticles and
extraneous postulates. I will show that the crucial physical interaction that
is at the basis of the remarkable and enigmatic features of the fractional
quantum Hall effect is gravity. The proof of this extraordinary connection is
straightforward and undeniable. Adding the factual gravity of the matter and
energy in the Universe as the dominant factor in the core Hamiltonian, as one
should, gives the result that \emph{the effective degeneracy of the Landau
levels is not only dependent on the applied magnetic field }$B$, \emph{as in
the existing theories, but also determined in a unique way by the carrier
density} $\rho$ \emph{in the 2D material}. This is the key point that
converges many aspects of the quantum Hall effects in a single unifying
reason. The gravitational energy of the electron in the factual Universe,
which amounts to about $-2\pi Gm_{e}\rho_{u}R_{H}^{2}$, is larger than
$10^{5}$ eV, where $\rho_{u}$ is the average density of the cosmic
matter-energy and $R_{H}$ is the Hubble size. This potential energy cannot be
renormalized away by subtracting a constant term because such a cosmological
term has no gauging freedom left; it is directly related to the observed
Hubble parameter. The cyclotron dynamics of the electrons in the 2D material
results in the induced relativistic gravitational potentials that act on the
mass, somewhat similar to the magnetic effects in electromagnetism. \emph{The
consequent gravitational modification of the quantum degeneracy from }%
$g_{m}=eB/h\equiv B/\phi_{0}$\emph{ to }$g=\left\vert B/\phi_{0}%
-2\rho\right\vert $\emph{ solves the primary puzzles of the fractional quantum
Hall effect, and also integrates it completely with the integer effect in a
single universal theory without quasiparticles}. This microscopic theory
reveals why the hypothetical schemes of fractionally charged quasiparticles
and the Composite Fermions with a hypothetical flux field were remarkably
successful as phenomenological descriptions, though far removed from the real
physical situation. The crucial point is that what is modified is the quantum
degeneracy alone, in the available \textquotedblleft action
space\textquotedblright, and not the effective magnetic field or the effective
charge of the particles. What could not be guessed earlier was that it is the
gravitational interaction with the cosmic matter that causes this
modification, while the Coulomb interaction and the degeneracy determine the
energy gaps in the lowest Landau level. This paper is written with the firm
insistence that the entire discussion will be based on what the relativistic
gravitational effect from the observationally cosmic matter would naturally
and robustly entails, without any additional assumptions about any
multi-particle effects in the material. Hence, I will not invoke any
quasiparticle or any hierarchical and nested schemes here. Much of these
findings are based on two mature developments, one on the dynamics in harmony
with the relativistic gravity of the matter-filled Universe \cite{Unni-Cosrel}%
, and the other on a new action mechanics that clarifies and integrates
classical and quantum mechanics on the robust basis of the Hamiltonian action
\cite{Unni-RQM}. However, this paper is self-contained, with all the relevant
derivations and proofs explicitly stated. The demonstrated success in
explaining both the integer and the fractional quantum effect in a seamless
and universal single particle theory without a single extraneous hypothesis
or conjecture, should be examined with interest.

\section{Introduction}

The Quantum Hall effect (QHE) was one of the most spectacular and
metrologically important discovery of the 20th century \cite{Klitzing-PRL}.
The phenomenon is the precise quantization of the Hall resistance as
$R_{H}=h/\nu e^{2}$ at the integer values of the `filling factor' $\nu=\left(
h/e\right)  \rho/B$, where $\rho$ is the carrier density in the 2D structured
semiconductor sample and $B$ is the applied magnetic field, or the magnetic
flux per unit area. The quantity $h/e$ is defined as the elementary `flux
quantum' $\phi_{0}$, and the filling factor is the ratio of the number density
(2D) of the carriers to the number density of the flux quanta, $\nu
=\rho/\left(  B/\phi_{0}\right)  $. Of course, expressing the flux density as
a number density in units of $h/e$ does not imply the physical discreteness of
the magnetic field, unlike the number density of the charge carriers. Yet, the
description is convenient and has a definite meaning in the quantum mechanics
of charged particles. The Klitzing constant at $\nu=1$, readily measurable
with an accuracy of a few parts in a billion, is now defined as the standard
of electrical resistance, $R_{/k}=h/e^{2}=25812.80745$ Ohms.
\begin{figure}
	\centering
	\includegraphics[width=0.85\linewidth]{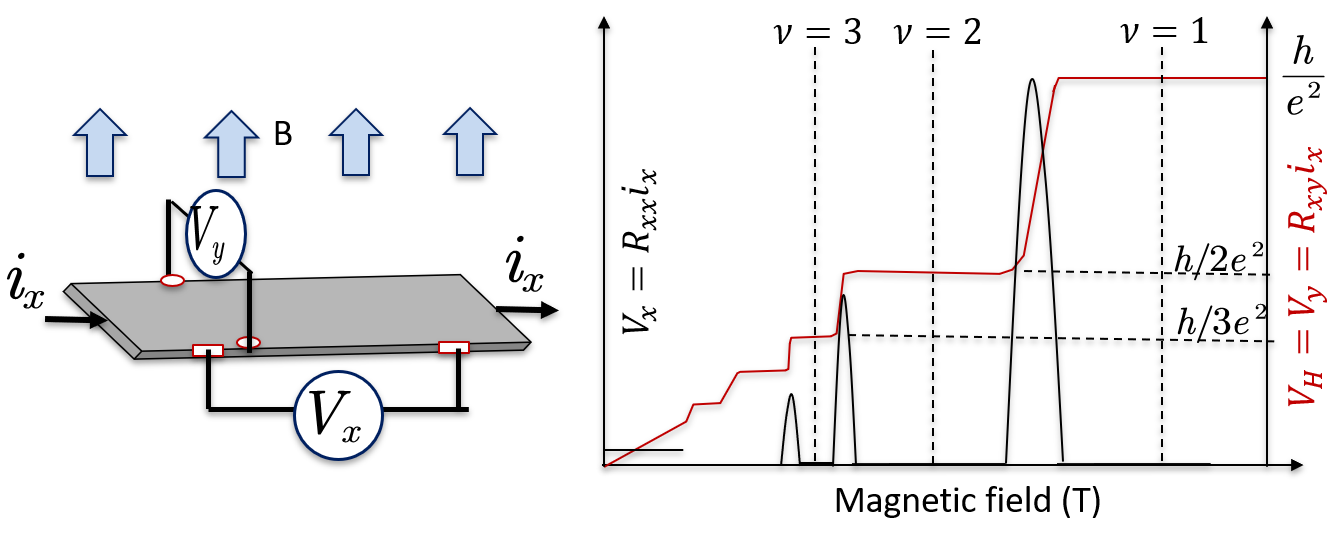}
	\caption{Left: The general scheme for the measurement of the Hall voltage and
			the longitudinal resistance of a 2D sample. A constant controlled current
			$i_{x}$ is passed through the sample kept in a uniform magnetic field $B$,
			applied perpendicular to the surface. The transverse voltage $V_{y}$ and
			longitudinal voltage $V_{x}$ are measured. The ratio $V_{y}/i_{x}=R_{xy}$ is
			defined as the Hall `resistance'. $V_{x}/i_{x}=R_{xx}$ is the longitudinal
			resistance. Right: The integer quantum Hall effect. $R_{xy}$ is quantized as
			$h/ve^{2}$ for the integer values of $\nu$, and $R_{xx}=0$ except at the
			transition jumps.}
	\label{fig:qhe}
\end{figure}

Apart from the quantization of the Hall resistance at the integer values of
$\nu$ as $R_{H}=h/\nu e^{2}$, the QHE state has the remarkable feature of the
dissipationless longitudinal current. Then the longitudinal resistance
$R_{xx}=0$. The energy levels of the electrons in the strong magnetic field,
called the Landau levels (LL), are quantized with the uniform quantization gap of
$E=\hbar\omega_{c}=\hbar eB/m.$ The cyclotron frequency of the closed orbital
motion of electrons in the 2-dimensional planar confinement region (the 2D
electron gas -- 2DEG) is $\omega_{c}$, and the energy of the levels is given
by $E_{n}=\left(  n+1/2\right)  \hbar\omega_{c}.$ The quantum degeneracy of
each of these levels is known to be equal to the number of the flux quanta;
the degeneracy per unit area is then the density of the flux quanta,
$eB/h=B/\phi_{0}$. The theoretical understanding of the precise quantization
of the Hall resistance is apparently simple and can be directly traced to the
Landau level quantization. \emph{When the carrier density is an integer
multiple of the flux quanta density, these levels are exactly filled to the
degeneracy} and the Hall resistance in the 2D system becomes
\begin{equation}
R_{H}=\frac{V_{y}}{i_{x}}=B/\rho e=\frac{B}{\nu\left(  eB/h\right)  e}%
=\frac{h}{\nu e^{2}}%
\end{equation}
The miracle is that the Hall resistance that depends on the ratio of the
magnetic field and carrier density has become independent of these physical
parameters, because the ratio is hidden as a pure integer. The real intriguing
feature in the observed data was the persistence of the quantized Hall
resistance as \emph{a plateau around the central integer values of the filling
factor} (figure \ref{fig:qhe}). The full explanation of these wide plateaux in
resistance is not simple. One needs to invoke the role of impurities in
localizing the electronic states and stabilizing the QHE states over a finite
width of the parameter space, leading to the plateaux over a considerable
range of the magnetic field \cite{Laughlin-1983}. However, \emph{the theory
is still in terms of the independent single particle picture, with the
interactions playing a minor role. Crucially, no multi-particle effects or
strong correlations are invoked} \cite{QHE-book}.

\begin{figure}
	\centering
	\includegraphics[width=0.7\linewidth]{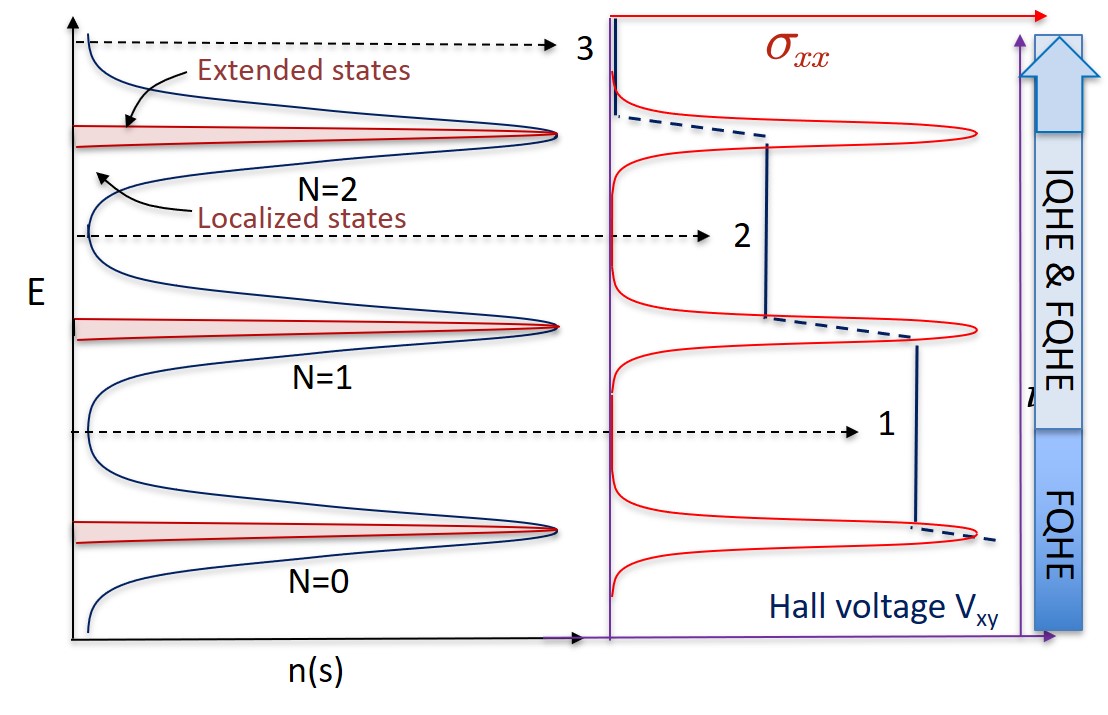}
	\caption{Summary of the current understanding of the integer quantum Hall effect. Left: The broadened Landau energy levels at a fixed magnetic field. The quantum states at the centre of each level are `extended' and participate in the conduction. The rest of the states are `localized', due to the presence of impurities. Right: The Hall voltage (blue) and the theoretical longitudinal conductivity $\sigma_{xx}$ (red) as the chemical potential of the electrons are scanned across the Landau levels, by increasing the number density $\rho$ and the filling factor ($\nu=h\rho/eB$) of the electrons.}
	\label{fig:hall-summary}
\end{figure}

Briefly stated, the QHE is related to the peculiar structure of the discrete,
yet broadened, Landau energy levels; all the states except the ones in a
narrow band of width $\delta$ at the centre, at the energy $E_{c}^{N}$ of each
level `$N$', are quantum mechanically localized due to the presence of a small
level of impurities (figure \ref{fig:hall-summary}). Therefore, as the levels are
filled, when the chemical potential $\mu$ is in the region of the localized
states, the Hall voltage remains a constant and the longitudinal conductivity
$\sigma_{xx}$ vanishes. The `Hall conductivity' is given by $\sigma_{xy}=\nu
e^{2}/h$. When the filling reaches the extended states (delocalized states),
$\sigma_{xx}$ becomes large for $\mu=E_{c}^{N}\pm\delta$, until it drops back
to zero, as the extended states are passed to reach the localized states in
the next level, where the Hall voltage changes to the next quantized constant
value. This scenario more or less fitted the early observations of the IQHE.
But, as more experiments measured the effect in better (high mobility)
samples, at higher magnetic fields (or lower carrier densities) and lower
temperatures, the entire landscape changed. Even in the case of the integer
effect at $\nu>1$, the behaviour of $\sigma_{xx}$ is more complicated.

\begin{figure}
	\centering
	\includegraphics[width=0.7\linewidth]{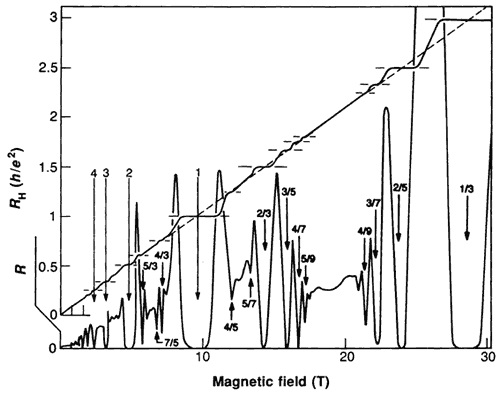}
	\caption{The `datascape' of the quantum Hall Effects \cite{Willett-1987}. Both the Hall resistance (upper trace) and the longitudinal resistivity (with the filling fractions indicated) are plotted. The wide stepped plateaux at various filling fractions are the quantized Hall resistances.}
	\label{fig:qhe-scape}
\end{figure}

Since the `Integer Quantum Hall Effect' (IQHE) happens when the charge density
is an integer multiple of the flux quantum density, the effect and the
appearance of the plateaux should end after $\nu=1,$ the smallest integer,
beyond which the charge density remains less than the degeneracy given by the
flux quantum density. Then the lowest Landau level is always partially empty. \emph{The surprise was the appearance of a very similar
quantization phenomenon for several fractional values of the filling factor}, $\nu<1$,
starting with the observation of the QHE at $\nu=1/3$. This presented a
perplexing challenge to the theory of the quantum Hall effect (figure
\ref{fig:qhe-scape}) \cite{FQHE-discovery}. How can the particles in an unfilled Landau level mimic a quantum phenomenon that can happen only in a fully filled level?  Some consensus on the theoretical understanding was possible
only by drastically deviating from the theory of the integer valued QHE, and
the construction of an entirely different kind of multi-particle theory
\cite{Laughlin-Nobel}. Subsequently, a proliferation of
fractions at which the Hall resistivity plateaux occur were observed, for both
$\nu<1$ and $\nu>1$, along with the vanishing of the longitudinal resistivity
\cite{FQHE-rev1}.

The theory of the IQHE remains intact as an independent particle
(non-interacting) theory, where the Coulomb interaction is involved in a minor
role in the explanation of the phenomenon. In contrast, the `Fractional QHE'
(FQHE), a very similar phenomenon in appearance at many fractional values of
the filling factor, like $1/3$, $2/5,2/3,..$ etc., posed an entirely new set
of theoretical challenges, with puzzling features remaining till today. There
has been remarkable progress in the phenomenological understanding of the FQHE
and the competing theoretical pictures can claim to be \emph{semi-microscopic
descriptions}. Quoting from the Nobel lecture of R. B. Laughlin
\cite{Laughlin-Nobel},

\begin{quote}
The fractional quantum Hall state is \textit{not} adiabatically deformable to
any non-interacting electron state...with a full understanding of the integral
quantum Hall effect there is no other possible conclusion. The Hall
conductance would necessarily be quantized to an integer because it is
conserved by the adiabatic map and is an integer in the non-interacting limit
by virtue of gauge invariance and the discreteness of the electron charge. So,
the fractional quantum Hall state is something unprecedented -- a new state of matter.
\end{quote}

It is easy to sense the possible routes to an effective theory of the FQHE,
exploiting a theoretical scheme of \textquotedblleft
quasiparticles\textquotedblright, given the fact that the parameters that
decide the quantized resistance $R_{H}=B/\rho e$ of the IQHE are not many.
Since the determining condition for the QHE is the exact filling of the
degeneracy, $\nu=\rho h/eB$ with $\nu$ as an integer, we may postulate a new
type of quasiparticles with an effective \emph{fractional charge} to model the
FQHE as a type of IQHE of such quasiparticles. For example, at $\nu=1/3,$ we
can invoke quasiparticles with the charge $e^{\ast}=1/3$ to get the effective
integer filling factor as $\nu^{\ast}=\nu/e^{\ast}.$ Another possibility is to
\emph{postulate an effective reduced magnetic field }$B^{\ast}(\nu)$. This
should be done in consistency with the key feature in the experimental data
that there is no QHE at an even fraction like $\nu=1/2$. Both these routes
have been taken, with the necessary sophistication and details to explain the
numerous fractions \cite{FQHE-rev1,Girvin-SP}. However, such theories have to
invoke a complicated hierarchy of physical phenomena, based on strong
multi-particle correlations and multiple adaptive hypotheses. There are
several unsatisfactory features that justify the belief that the correct
picture is still very much veiled.

The key assertion in this article, with the required proofs and demonstration,
is that \emph{there is an entirely new and clearly present physical factor
that has been overlooked so far in the theory of the quantum Hall effects}.
The strong reasons that motivated the need for a new theory of the QHE are to
be spelt out before we discuss the details of the missing physics of gravity,
because any suggestion that gravity plays a key role in a low temperature
condensed matter phenomenon would understandably be taken as unbelievable, and
even ridiculous.  At the same time, we should keep in mind that we still lack a
convincing microscopic theory of the FQHE, in spite of the remarkable success
of the quasiparticle theories, especially the Composite Fermion (CF) theory
\cite{Jain-book}. The fact that the gravitational interaction energy of the electrons with the matter in the Universe exceeds all other contributions to the free energy of the electron has remarkable consequences to electron's dynamics in several physical situations \cite{Unni-Cosrel}.

\section{The Need for a New Microscopic Theory}

Why do we consider the present theories of the FQHE as phenomenological and
seek a more fundamental microscopic theory? The primary reasons are two: 1)
Both the IQHE and the FQHE happen seamlessly in the same sample, sequentially
and alternating, as either the magnetic field is varied or the carrier density
is varied. The phenomenological equivalence is particularly striking for the filling factors larger than $\nu=1$, where the IQHE occurs. This suggests the plausible universality of the underlying physics,
rather than repeated transitions between the simple single particle behaviour
and the complex multi-particle phenomena, in the same sample in identical
experimental conditions, as one parameter is varied. Given a graphical
representation of the data, the FQHE is very similar to the IQHE, and the two
occur at the vicinity of each other, alternating. 2) While examining many
physical situations that involve noninertial dynamics and closed orbitals in
quantum mechanics, in the enormous relativistic gravitational potentials of
the entire matter and energy in the Universe, I have found significant
physical effects that have an immediate and direct consequence to the
physics of the quantum Hall effects. \emph{Since the cosmic matter at the
average density of about }$10^{-26}$\emph{ kg/m}$^{3}$\emph{ is
observationally verified, its immense gravity is an undeniable reality that
one should include in all relevant physical situations, as confirmed by
several concrete experimental evidences} \cite{Unni-Cosrel,Unni-ISI,Unni-TPU}. That is, any statement that dismisses the reality and enormity of cosmic gravity is an unjustified denial of the cosmic matter and energy that have been observationally verified.

The first point is readily noticed and it is widely discussed. Quoting from
Laughlin's Nobel lecture,

\begin{quote}
So the fractional quantum Hall state is something unprecedented -- a new state
of matter. Its phenomenology, however, is the same as that of the integral
quantum Hall state in almost every detail (Tsui et al., 1982). There is a
plateau. The Hall conductance in the plateau is accurately a pure number times
$e^{2}/h$. The parallel resistance and conductance are both zero in the
plateau. Finite-temperature deviations from exact quantization are activated
or obey the Mott variable-range hopping law, depending on the temperature. The
only qualitative difference between the two effects is the quantum of Hall conductance.
\end{quote}

Yet, the fractional effect required an entirely different physical premise and
theoretical formulation. Early attempts discussed the possibilities of
multi-particle effects due to the strong Coulomb interaction, phase
transitions, and fractionally charged quasiparticles. J. K. Jain showed that
the FQHE can be accessed from the IQHE by adding an even number of
\textit{hypothetical flux quanta of a gauge field} (not the familiar
electromagnetic field) to each electron, generating new quasiparticles, or
more precisely, particle-vortex composites, preserving their fermionic
character \cite{Jain-PRL}. This was a scheme for obtaining a reduced effective
field $B^{\ast}$ in the sample, instead of the applied magnetic field $B$;
adding an even number of the `new kind of flux quantum' for every particle
ensured both the fermionic character of the quasiparticles and a specific
dependence of $B^{\ast}$ on the filling factor $\nu$, which was crucial to get
the quantized Hall effect at the filling factors $\nu=1/3,1/5$, etc. Also,
this scheme guaranteed the absence of the QHE at $\nu=1/2$, corresponding to
the half-filling on the Landau level. Quoting from Jain's paper that
introduced the `Composite Fermion' approach to the FQHE \cite{Jain-PRL},

\begin{quote}
While the integer QHE (IQHE) is thought of essentially as a non-interacting
electron phenomenon, the fractional QHE (FQHE) is believed to arise from a
condensation of the two-dimensional (2D) electrons into a \textquotedblleft
new collective state of matter\textquotedblright as a result of
inter-electron interactions. Even within the FQHE the \textquotedblleft
fundamental\textquotedblright fractions 1/3, 1/5, . . . play a special role
and the other fractions are obtained in a hierarchical scheme in which a
daughter state is obtained at each step from a condensation of the
quasiparticles of the parent state into a correlated low-energy state.
\end{quote}

One is justified in thinking that it is unnatural that the physical system
repeatedly alternates between a single particle phenomenon, with interactions
playing a minor role, and a complex multiparticle phenomenon, where the
Coulomb interaction becomes the dominant factor, as one of the parameters is
smoothly varied. This happens many times when $\nu>1$. The drastic
transformation of the physical description between the two alternatives is
required repeatedly, while the basic nature of the phenomenon remains exactly
the same. Also, the FQHE involves not one, but many phenomena, depending on
the nature of the fractional filling factor (the hierarchy mentioned by Jain).
The QHE plateaux at the various filling fractions $\nu=h\rho/eB$ can be
obtained by either increasing the field $B$ or \emph{by decreasing the carrier
density }$\rho$. Since these samples are field effect transistors, the amount
of charge stored when the gate voltage is varied is like in a capacitor,
$Q=CV$. Thus, a voltage sweep is equivalent to the charge density sweep. For
$\nu<1$, the plateaux are obtained by reducing the density of particles from
its value at $\nu=1$. But, reducing the density increases the average
interparticle separation and reduces the average interaction energy. \emph{It
is then certainly unnatural to resort to multi-particle effects and a dominant
Coulomb interaction to take over the physical situation,  from the single
particle `non-interacting' behaviour, as the particle density is reduced and
the interparticle separation is increased.}

\begin{figure}
	\centering
	\includegraphics[width=0.9\linewidth]{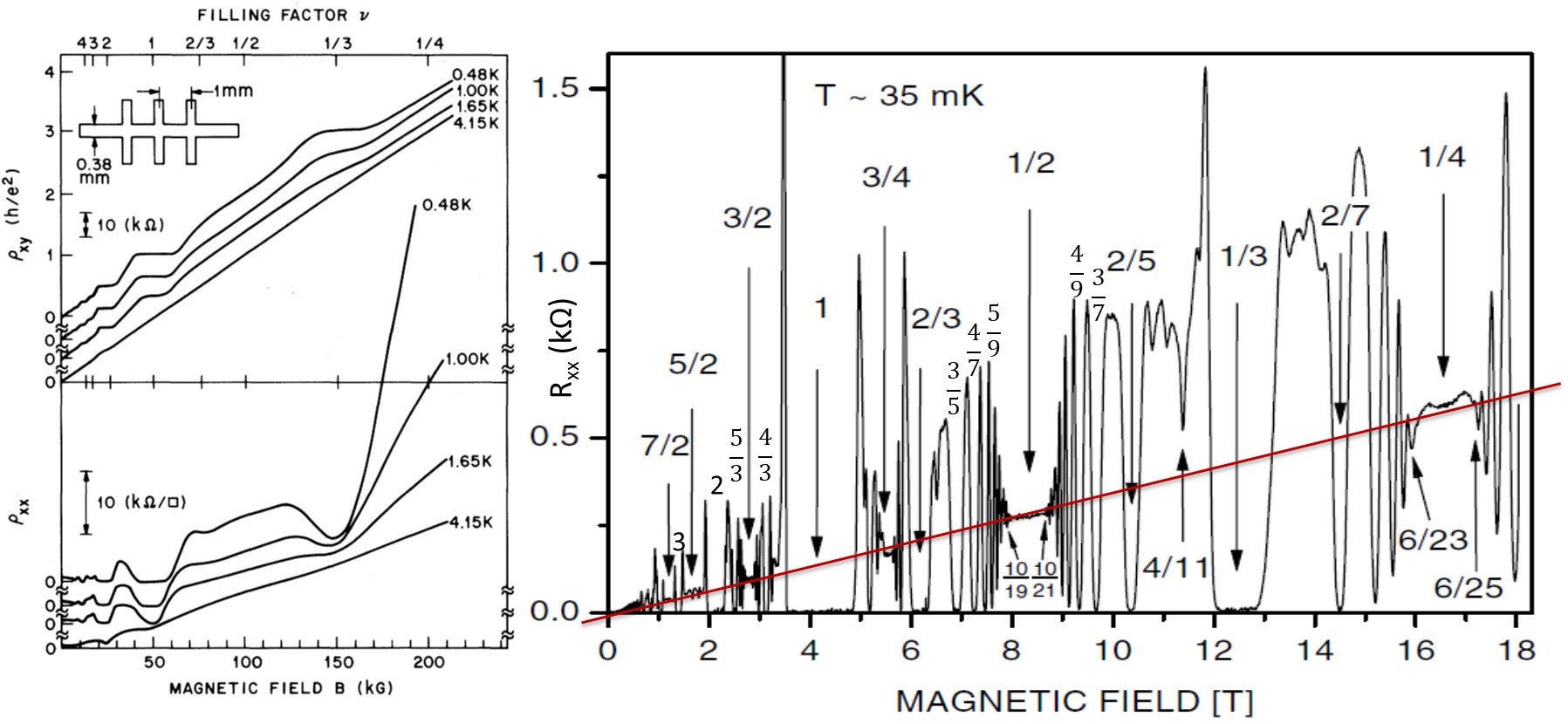}
	\caption{Left: The first observation of the FQHE at 1/3 \cite{FQHE-discovery}. This data indicate that the magnetic field experienced by the electrons at $\nu=1/2$ is the applied field $B$, and not the diminished field $B^{\ast }=B(1-2\nu)$. Right: The plot highlighting the vanishing longitudinal resistance at several fractional filling factors \cite{Stormer-data}. It is clear that the magnetic field responsible for the background magnetoresistance (linear in the applied field) is the factual applied field that can be extrapolated from the $R_{xx}$ at very low field (red line).}
	\label{fig:longir}
\end{figure}

The Hall resistivity is proportional to the applied magnetic field, with the
same proportionality factor $1/\rho e$ as in the classical Hall effect, for
those values of the filling factor where there are no plateaux. \emph{Even on
the plateaux, the mean Hall resistivity follows the classical value} (see the
figure \ref{fig:longir}:left panel). If the FQHE arises from composite particles
that can internally modify the applied magnetic field, one has to invoke one
value of the residual field for the quantum degeneracy estimate and another
for the Hall effect. Or, one may have to assign different roles to different
sets of charge carriers. Even more revealing is the behaviour of the
longitudinal magnetoresistance near the filling factors of even fractions
$1/2$ and $1/4$. The longitudinal resistance is proportional to the applied
magnetic field at the filling factors where there is no QHE, with the same
proportionality constant one measures at the low magnetic field (indicated
with a red line in the figure \ref{fig:longir}). If the formation of the Composite
Fermions actually reduced the effective magnetic field experienced by the
charged carriers, one does not expect this behaviour of the longitudinal
magnetoresistance, especially near the even fractions.

Finally, both the IQHE and the FQHE are observed in the 2D system graphene
\cite{Graphene-rev}. Monolayer graphene has special conduction and valence
bands that touch, with a linear dispersion relation, $E\propto p$. The
quantization of energy in the presence of a strong magnetic field is now
$E_{n}=\pm v_{F}\sqrt{2\hbar eBn}$, instead of the cyclotron spectrum
$E_{n}=\left(  n+1/2\right)  \hbar\omega_{c}$ for the 2D electron gas in the
structured 2D materials. $v_{F}$ is the Fermi velocity determined by the
inter-site hopping frequency and the lattice spacing. The IQHE is seen in both
the electron branch and the hole branch in graphene, symmetrically. Graphene
entered the experimental scene much after the quasiparticle theories of the
QHE were developed, after a monolayer of graphene was realized in 2004
\cite{Novoselov-QHE}. Both the integer QHE and the fractional QHE were
observed in graphene at very low carrier densities of about $10^{11}/cm^{2}$.
I think that the QHE in graphene poses some hard challenges to the
microscopic theory of the regular quantized plateaux of the Hall resistance.
While the electron-hole symmetry in the system is near perfect, the fact
remains that the hole is already a `quasiparticle'. It is the single particle
representation of a multiparticle transport effect, where the valence
electrons nearly fill the band. The identity of the electrons that actually
execute the dynamics keeps changing -- what is in transport is the `absence'
of a negatively charged particle. This means that the picture of `flux tubes
attaching to the absence of a particle' becomes further removed from reality.
It is not easy to grasp the symmetries in the data in a scheme where
multi-particle collective effects create new quasi-particles in complex
patterns and then they exhibit the QHE, because the environment of the
electrons and the holes are entirely different. The Coulomb dominated
multi-particle effects cannot be identical for electrons and holes in any
natural description of the phenomenon of the QHE. More recent measurements in
the axially symmetric Corbino geometry \cite{Dogopolov,Zeng} definitely prompt
a second closer look at even the theory of the IQHE
\cite{Klitzing-Physica,Klitzing-Weis}.

\subsection*{Summary of the reasons}

The dissatisfaction with the impressive current theories of the QHE is not
about their efficacy or their phenomenological success, but about their
naturalness, physicality and the inclusiveness of relevant physics. Simplicity
might be another desired quality, but one has to admit that complicated
phenomena can have complicated details in theories. It is useful to summarize
the points that motivate one to go beyond the present theories.

\begin{enumerate}
\item The IQHE and the FQHE occur seamlessly alternating in the same sample as
either the charge density or the magnetic field is varied. The theories invoke
entirely different physical mechanisms for very similar phenomena, in
different parts of the QHE skyline, alternating repeatedly.

\item The IQHE is described by essentially a non-interacting single particle
theory, whereas the FQHE in the neighbourhood of the same parameters (magnetic
field and carrier density) requires a strongly interacting multi-particle
theory with quasiparticles and condensations. In fact a hierarchy of
multi-particle effects is postulated. There is a problem of naturalness.

\item There is not a unique theory of the FQHE, but there are two or more
non-equivalent theories with entirely different physical basis
\cite{Jain-IndianJl,Jain-PRL,Haldane-FQHE,Halperin-FQHE}. They have common as
well as widely differing characteristics.

\item The mean Hall resistivity is proportional to the applied magnetic field,
$R_{H}=B/\rho e$, as for the classical Hall effect. The longitudinal
magnetoresistance $R_{xx}$ in regions without the QHE remains faithfully
proportional to the applied magnetic field, including at the even fractional
filling factors like 1/2 or 1/4. (It has the approximate general form
$R_{xx}=\alpha B\left(  dR_{xy}/dB\right)  $, and when not on a QHE plateau in
the $\nu<1$ region, $R_{xx}\approx\alpha B/\rho e$). But the Composite Fermion
(CF) theory in particular makes use of a internally diminished magnetic field
to derive the FQHE as the IQHE of CF quasiparticles. In the CF theory, the
internal effective field goes to zero at these even fractions. So, there is a
genuine conflict with the experimental evidence. Further, with $\nu=1/2$ as
the zero-point of the `internal magnetic field' experienced by the particles,
$B^{\ast}=B\left(  1-2\nu\right)  $, the residual field $B^{\ast}$ becomes
negative, in the direction opposite to the applied field, when $\nu>1/2$
(figure \ref{fig:cf-field}). Then, the sign of the Hall voltage should reverse,
which it does not. The only way out in the CF picture is to introduce another
hypothesis, of positively charged hole counterparts taking over the dynamics,
instead of the electrons, which is not credible.

\item The QHE in graphene shows remarkable symmetry in the electron and hole
branches. This is more natural in an independent particle view because holes
are already quasiparticles from a multiparticle transport picture, and their
momentum space environment is different from that of the conduction electrons.

\item The fractions with $\nu>1$ require nested hierarchical models in the
present theories.

\item As for the FQHE theory with fractionally charged quasiparticles, the
proliferation of fractions renders the theoretical picture complicated and unmanageable.

\item Finally, we want to stress a very important general reason. \emph{All
physical effects in nature, irrespective of their mathematical descriptions,
should be traceable to one of the four interactions we know of.} In
particular, all bulk condensed matter phenomena should be traceable to the
only two long range fundamental interactions in nature -- electromagnetism and
gravity. Therefore, the present CF theoretical picture of the FQHE that uses a
hypothetical gauge field can only be an effective theory, and not the true
microscopic theory.

\begin{figure}
	\centering
	\includegraphics[width=0.65\linewidth]{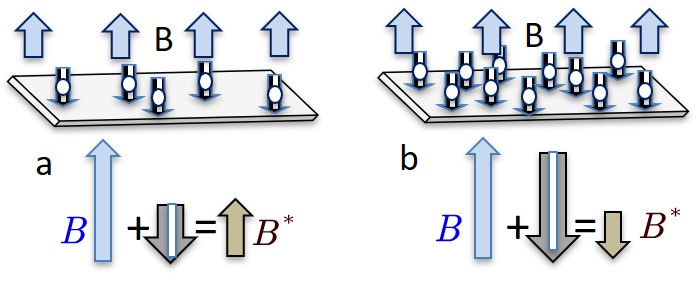}
	\caption{The CF theory relies on the reduction of the effective magnetic field $B^{\ast}$ experienced by the charged particle, by adding a fictitious field of the even number of flux quanta associated with the Composite Fermions to the applied field $B$. $B^{\ast}=B\left( 1-2\nu\right) $, where $\nu =\rho/\left( B/\phi_{0}\right) $. a) For $2\nu<1$, the effective field is still in the direction of the applied field. b) At larger carrier density, when $2\nu>1$ as in the FQHE filling factor $\nu=2/3$, the effective field becomes opposite to the applied field. However, the Hall voltage does not change sign in the vicinity.}
	\label{fig:cf-field}
\end{figure}

\end{enumerate}

\begin{figure}
	\centering
	\includegraphics[width=0.7\linewidth]{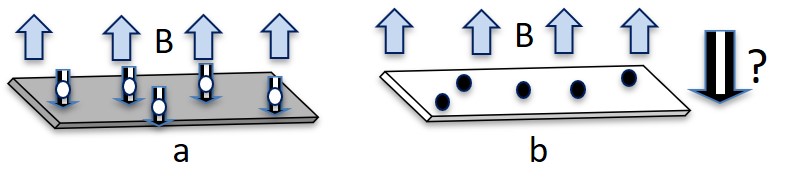}
	\caption{What exactly is the construction of a Composite Fermion from a hole
		carrier? a) The notion of an electronic CF, where each conduction electron in
		a Landau level is associated with an even number of the `quanta of a gauge
		field' that can physically diminish the total magnetic field seen by the
		electrons. b) Hole carriers are physically the dynamically mobile `absence of
		electrons' (dark blobs) in a uniform background of negative charges (shown as
		the white background), and not a physical particle. Then, the notion of a hole
		CF is not straightforward. How does one associate the even fluxtubes with the
		localized absence? }
	\label{fig:holecf}
\end{figure}

The points 4 and 5 are particularly important for the phenomenologically
successful CF theory initiated by J. K. Jain. Focusing specifically on the microscopic details of the QHE with hole conductivity, I see serious conceptual issues with making Composite
Fermions from an already `quasified' collective notion of a particle that we
call the `hole'. The CF postulate is the notion of the hypothetical `even fluxtubes'
associated with every conducting electron (figure \ref{fig:holecf}). However, this
scheme is difficult to carry through in detail with a hole, which is\emph{ the absence of an
electron} in a uniform occupation of electrons. Unlike the positron, there is
no real particle `in the hole', so to say. Now, a CF associated with a hole is
in some sense a `second quasification'; in the CF picture, the notion of the
flux quanta that get attached to the hole charge becomes a concept `twice
removed' in that \emph{both the `hole' charge and the flux quanta are
fictitious}. One needs to either associate the even fluxtubes with the charge
density of the negatively charged background except at the location of the
hole, to match the absence of the flux with the absence of the particle, or
resort to some even bizarre scheme, to get a CF out of a hole.

There might be other aspects that can be listed, depending on the theoretical
preferences. But, these are my primary reasons for the conviction that the
correct theory is yet to be formulated. Despite the impressive progress in the
theory of the FQHE, there is no unique theory on which there is a general
consensus. One may even say that what one is looking for is a much desired
coherent and entirely physical `Copernican view', at a time we have a fairly
successful `Ptolemaic view', but full of epicycles and their hierarchies. From
the abstract of J. K. Jain's `A note contrasting two microscopic theories of
the fractional quantum Hall effect' \cite{Jain-IndianJl},

\begin{quote}
Two microscopic theories have been proposed for the explanation of the
fractional quantum Hall effect, namely the Haldane--Halperin hierarchy theory
and the composite fermion theory. Contradictory statements have been made
regarding the relation between them, ranging from their being distinct to
their being completely equivalent. This article attempts to provide a
clarification of the issue. It is shown that the two theories postulate
distinct microscopic mechanisms for the origin of the fractional quantum Hall
effect, and make substantially different predictions that have allowed
experiments to distinguish between them.
\end{quote}

However, one common feature is the hierarchical and nested nature of the
theories, where multi-particles effects are postulated repeatedly when faced
with new series of fractions in the FQHE. What we will achieve in this paper
is a unified and coherent single particle theory of all the quantization
filling factors of both the IQHE and the FQHE. \emph{I will not make any new
postulates}. Instead, I will rely on verifiable physics and cosmology, based on facts. We
will be able to derive all the prominent fractions at which the QHE occurs for
$\nu\geq1/3$, with the natural explanation for the absence of the even
fractions in the FQHE. I do not claim to deal with all detailed aspects of the
quantum Hall effects, like the detailed finite temperature behaviour of the
longitudinal resistivity in the FQHE, which are intrinsically complicated and
still being investigated experimentally. I will mention those issues that
remain and need further detailed attention.

\section{A Serious Spin-Phase Puzzle}

Now we can explore the gravitational interaction between the electrons in the
laboratory and the entire matter and energy in the Universe. While there are
many ways of introducing this connection in a quantitatively accurate manner,
it might be more appropriate to use a physical example that involves the very
dynamics of the electron that is relevant for the QHE. A serious fundamental
puzzle arises in the quantum dynamics of a single electron in a large magnetic
field, which requires urgent resolution \cite{Unni-PhysNews}. 

\begin{figure}
	\centering
	\includegraphics[width=0.25\linewidth]{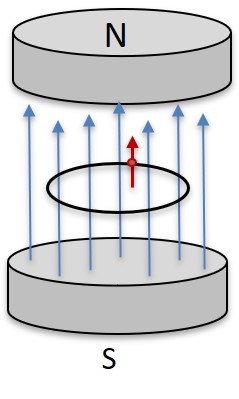}
	\caption{The simple scenario of the phase puzzle. The quantum phase of the
		electron in the electromagnetic field should be an integer times $2\pi$. But,
		the contribution from the interaction energy $\mu\cdot B$ of the magnetic
		moment in the field is $\pi$, violating this requirement.}
	\label{fig:phase-puzzle}
\end{figure}

Consider the simple quantum dynamics of a single electron in a uniform
magnetic field (figure \ref{fig:phase-puzzle}). A serious problem arises when there are closed cyclotron
orbitals, in a way that has not been noticed. The electron orbits in quantum
mechanics should obey the quantization condition of a cyclic phase of $2\pi$,
for the single valuedness of the wavefunction,

The quantum phase of the spinless charge in the magnetic field is
\begin{equation}
\delta\varphi=\frac{1}{\hbar}\oint p\cdot dx=\oint\left(  m\dot{x}-eA\right)
dx=2n\pi
\end{equation}
From this we say that the `flux enclosed is quantized', since the
electromagnetic contribution for an electron ($-e$) in the orbit is%
\begin{align}
\delta\varphi &  =\frac{1}{\hbar}\oint eAdx=\frac{1}{\hbar}\iint e\left(
\nabla\times A\right)  dS=\frac{2\pi e}{h}\phi_{B}=2n\pi\\
&  \rightarrow\frac{e}{h}\phi_{B}\equiv\frac{\phi_{B}}{\phi_{0}}=n
\end{align}
Here, $\phi_{B}$ is the magnetic flux and the ratio $h/e$ is defined as the
elementary flux quantum $\phi_{0}$. Factually, it is the phase of the
wavefunction that is quantized, which is the fundamental condition for closed
orbitals irrespective of the nature of the forces involved in the dynamics.
The quantization is very similar to the case of the superconductor where the
flux enclosed is quantized in units of $h/2e,$ because it is the `Cooper
pairs' that are the charge carriers.

However, this quantum condition is obviously altered because the spin and the
magnetic moment of the electron need to be considered for the quantum phase
(figure \ref{fig:phase-puzzle}). The interaction energy of the electron magnetic
moment with the magnetic field, $E=\mu\cdot B=g\mu_{B}\left(  s\cdot B\right)
$, causes an additional phase shift over the orbit with the period
$T=2\pi/\omega_{c},$%
\begin{equation}
\delta\varphi_{s}=\frac{1}{\hbar}\int Edt=\pm\frac{1}{\hbar}\int g\mu
_{B}sBdt=\pm\frac{eBT}{2m}=\frac{\pi eB}{m\omega_{c}}=\pm\pi
\end{equation}
The phase contributed by the polarized spin in the full orbital is only $\pi$,
instead of the required $2\pi$! This spoils the phase quantization and the
quantum condition, and hence the simple theory. \emph{This problem does not
arise in the case of superconductivity because two spin-1/2 particles are
involved with either 0 or 1 total spin.} Then the additional phase is 0 or
$2\pi$. In fact, this troublesome result could have been simply guessed from
our experience with superconductivity where the orbital phase of a pair of
electrons (due to the magnetic moment) is either 0 or $2\pi$; then, obviously,
the phase of a single electron can only be $\pi$. The mitigation of this
disastrous result requires finding \emph{another source of quantum phase} that
will provide the correct phase to satisfy the closure condition of $2\pi$. Now
I will show that the gravitational interaction with the matter in the Universe
at the critical density, as observed, contributes exactly the correct quantum
phase. \emph{This serves as the proof of the reality of gravitational
interaction, with the factually existing and the observationally confirmed
cosmic matter-energy}. Given the strangeness in novelty of the physics of the
gravitational consideration in the condensed matter problem, it might be
desirable, and indeed instructive, to briefly discuss an analogy for the
solution first.

What we need is to see whether there could be an unnoticed physical interaction that
could provide the missing `$\pi$' in the phase puzzle. Suppose there is a
large uniform cloud of charged particles with a very small number density, so
small that it remains undetected in any local direct measurement. If the spatial size
of the charge distribution is astronomically large, the electric fields are nearly zero, but the uniform
electric potential can be large and significant. Ignoring the `edge effects', we may
estimate it as
\begin{equation}
\phi_{e}\approx\frac{1}{4\pi\varepsilon_{0}}\int\limits_{0}^{R}\frac{4\pi
r^{2}e\rho dr}{r}=\frac{e\rho R^{2}}{2\varepsilon_{0}}%
\end{equation}
Any relative motion will result in a vector potential, $A\approx v\phi_{e}/c$,
and the cyclotron motion involving a rotation will generate an induced
magnetic field $B_{i}=\nabla\times A/c=2\phi_{e}\omega_{c}/c^{2}$. This will
couple to the magnetic moment as $\mu\cdot B_i$. The resulting extra quantum phase will be
\begin{align}
\delta\varphi &  =\frac{1}{\hbar}\int\mu_{e}\cdot B_{i}dt=\pm\frac{1}%
{\hbar\varepsilon_{0}c^{2}}g\mu_{B}se\rho R^{2}\omega_{c}T\\
&  =\pm\frac{1}{2m\varepsilon_{0}c^{2}}e^{2}\rho R^{2}\omega_{c}T=\pm\pi
\frac{e^{2}\rho R^{2}}{m\varepsilon_{0}c^{2}}%
\end{align}
Clearly, the required closure phase can come from this induced field if
$e^{2}R^{2}\rho/\varepsilon_{0}mc^{2}\approx1$. It is quite remarkable that a
number density of charges as small as $10^{-14}/$m$^{3}$ is enough to generate
this large phase shift, if the charge cloud extends over the solar system. If
it is as large as the galaxy, a charge density as small as $10^{-27}e/$m$^{3}$
would suffice. But, it is
even more remarkable that there is not even that much of sustained charge
asymmetry around us; the Universe is by and large charge neutral, electrically
\cite{UG-charge}. Therefore, the missing phase cannot come from the
electromagnetic interaction.  \emph{However, the Universe is not charge
neutral, gravitationally}! The mass and energy are the charges of gravity and
there is an enormous amount of matter in the vast Universe, at the average
mass density of $10^{-26}$ kg/m$^{3}$, equivalent to a few Hydrogen atoms per cubic meter. The physics of relativistic gravity is very similar to the physics of electromagnetism when the nonlinear effects
of gravity are small. The gravitational potential follows a very similar
expression, with the size $R$ of the order of the Hubble radius. We will
explore the straightforward consequence to the cyclotron dynamics in the large
gravitational potential of the cosmic matter. Unlike the case of the particle
with the spin, where the gravitational effect involves both a quantum phase
and the strong gyroscopic torque from the difference in the interaction energy
for the two directions of the spinor spin, a spinless massive particle in the
induced relativistic potentials just experiences a dynamical phase and a
modification of its dynamical action, which are of crucial importance in the
physics of the quantum Hall effects.

\section{New Physics Input -- Gravity}

The main physical input for the new microscopic theory of the QHE is cosmic gravity.
From the physical point of view of fundamental interactions,
the \emph{charge carriers in the condensed matter have a mass, in addition to
the electric charge, and the mass is also the `gravitational charge'}. In
addition, there is the quantum spin, along with the magnetic moment. The spin
of the particle couples to the gravitational field generated by the relative
currents of masses (gravitomagnetic field) in a way similar to the coupling of
the magnetic moment with the magnetic field \cite{Wheeler-book}. The phase of
the wavefunction changes by coupling to the electromagnetic potentials as well
as the gravitational potentials. The last thing one would expect in the
microscopic physical description of a condensed matter phenomenon is the
gravitational influence as a primary effect. The role of gravity in the
relevant Hamiltonians is usually assumed to be negligible, since the
contribution from the largest local source -- the Earth -- is indeed
negligible. But, this prejudice ignores the fact that the entire matter in the
Universe gravitationally interacts with every electron in a piece of matter.
While the cosmic gravitational field and the gravitational curvature are
nearly zero, the gravitational potential is not, and that is what is relevant
for the quantum dynamics and its Hamiltonians. In fact, the inclusion of the
interaction Hamiltonian $H_{g}=-m\Phi_{u}$ in the Schr\"{o}dinger equation,
where $\Phi_{u}$ is the gravitational potential due to all the matter in the
Universe, is necessary in every case because \textit{the actual gravitational
potential is numerically of the order of }$c^{2}$! Thus, \textit{it is the
largest term in any physical Hamiltonian}, but often playing a constant
passive role. This serious omission of the large background gravity of the
cosmic matter in our fundamental theories happened because when these theories
(relativity, quantum mechanics etc.) were completed no significant knowledge
about the matter content and the extent of the Universe existed. Modern
cosmology developed and matured much later, after 1930. Besides, in most
nonrelativistic inertial phenomena, the large gravitational potential has only
the effect of a common constant, which is not detectable. However, \emph{in
situations of acceleration and rotation of the particles, the relativistic
gravitational potentials are not only significant, but play a dominant role in
the quantum dynamics} \cite{Unni-ISI,Unni-TPU}. This is the new essential
input in our theory of the QHE. This input of the gravity of the cosmic matter
is an unavoidable necessity, because the matter-energy in the vast Universe is
real and factually present.

\subsection{The cosmic gravitomagnetic potentials in dynamics}

The basic dynamics of charged particles in a constant magnetic field is the
cyclotron motion in the perpendicular plane, with the Lorentz force balancing
the centrifugal force. This is a superposition of two harmonic motions in the
x and y directions, with the relative phase $\pi/2$. The quantum mechanics of
this motion leads to the Landau levels (LL) with equally spaced energy levels that
are degenerate.
\begin{align}
\frac{evB}{m^{\ast}}  &  =v^{2}/r\\
E_{n}  &  =\hbar\omega_{c}(n+1/2);\quad\omega_{c}=eB/m^{\ast}%
\end{align}
Here, the mass is the effective mass in the solid. The degeneracy of each
LL, per unit sample area, is equal to the $B/\phi_{0},$ where
$\phi_{0}=h/e$ (I will derive this result later).

\begin{figure}
	\centering
	\includegraphics[width=0.8\linewidth]{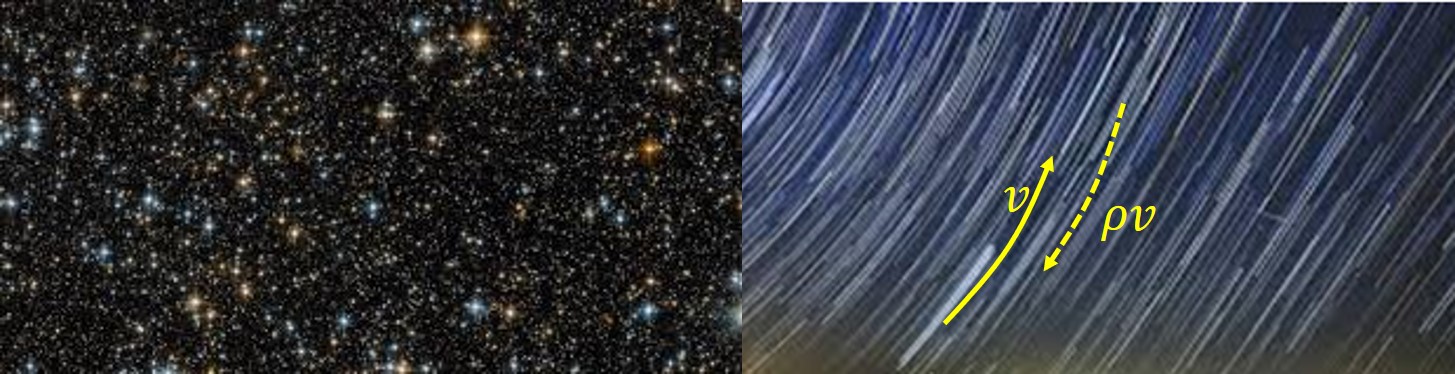}
	\caption{In the matter-filled universe, the relative motion at the velocity
		$v$ causes a relative matter current density $\rho v$ and its relativistic
		gravitational potentials. Since the static gravitational potential $\Phi_{u}$
		is very large, the motion-induced gravitational effects are also very large.}
	\label{fig:mass-current}
\end{figure}

\begin{figure}
	\centering
	\includegraphics[width=0.5\linewidth]{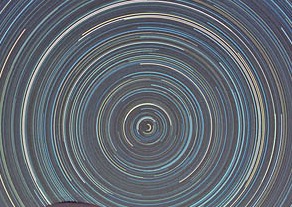}
	\caption{Cosmic matter in a rotating frame. The curl of the gravitational potential of this current is locally $-2\Omega\Phi_{u}/c^{2}=-2\Omega$.}
	\label{fig:star-curl}
\end{figure}

\emph{This dynamics happens in the presence of the vast amount of
matter-energy in the Universe and its gravitational potential, and not in
empty space as usually assumed}, as we already stressed. The cosmic
gravitational potential $\Phi_{u}$ is approximately about $\int{4\pi G\rho
rdr}$, where the integration extends to the Hubble radius. A numerical
estimate of this potential in the average rest frame is about $10^{17}%
m^{2}/s^{2}$, which is numerically close to the square of the velocity of
light, $c^{2}$. This is in the frame that is at rest (comoving) relative to
the frame of the average matter distribution in the Universe. In a slowly
moving frame at velocity $v_{i}$, like that of a particle in a current, there
is the extra relativistic gravitational potential as well, akin to a vector
potential. This is given by
\begin{equation}
A_{i}=\left(  \frac{v_{i}}{c}\right)  \Phi_{u}%
\end{equation}
For more exact results, we will work with the metric of the real Universe, the
spatially flat FLRW metric. In a nearly isotropic and homogenous Universe,
symmetry dictates that the physical metric is
\begin{equation}
ds^{2}=-c^{2}dt^{2}+a^{2}(t)(\frac{dr^{2}}{1-kr^{2}}+r^{2}d\Sigma^{2})
\end{equation}
Since it is known from observations that the average density of the Universe
is close to the critical density, the curvature parameter $k$ is nearly zero.
Also, the time dependent scale factor $a(t)$ changes very slowly
($a(t)\simeq1+H_{0}t$, where $H_{0}=10^{-18}\,m/s/m$) and can be considered
nearly a constant over time scales relevant to the whole history of physics,
or even civilizations. Thus for the observer to whom the Universe is isotropic
and homogenous,
\begin{equation}
ds^{2}\simeq-c^{2}dt^{2}+a^{2}(dr^{2}+r^{2}d\Sigma^{2})=-c^{2}dt^{2}%
+(dx^{2}+dy^{2}+dz^{2})
\end{equation}
where the nearly constant factor $a$ is absorbed into the coordinates. Cosmic
evolution has a universal time, as indicated in its gravitational metric.
Transforming from $(x,ct)$ of the privileged cosmic frame to $(x^{1}%
,x^{0}=x^{\prime},ct^{\prime})$ of a moving frame, representing the physical
motion relative to the cosmic matter,
\begin{equation}
x^{\prime}=x-\left(  \frac{v}{c}\right)  ct,\quad t^{\prime}=t
\end{equation}
we get the new metric (showing only components that are relativistically
modified)%
\begin{equation}
g_{ik}^{\prime}=%
\begin{bmatrix}
-\left(  1-v^{2}/c^{2}\right)  & -v/c\\
-v/c & 1
\end{bmatrix}
\end{equation}
Other components are $\delta_{ik}^{\prime}$. The new non-zero metric
components in the moving frame are $g_{00}^{\prime}=-(1-v^{2}/c^{2}%
),~g_{01}^{\prime}=-v/c,$ and $g_{ii}^{\prime}=1$. The remarkable fact of the Lorentz factor emerging from the Galilean transformations, and its implications, are discussed elsewhere \cite{Unni-Cosrel}. For the uniform motion, the relevant velocity dependent (magnetic) potential  is
$g_{0i}=-v_{i}/c$, and this is exactly like the gauge potential in
electrodynamics (figure \ref{fig:mass-current}). The gravitational `vector' potential $\vec{A}_{g}\equiv
cg_{0i}=-\vec{v}$ couples to the gravitational mass as $mA_{g}$. In inertial
motion, this is unobservable, but on nonlinear dynamics this coupling has
physical effects through non-zero $dA_{i}/dt$ and $Curl(A_{i})$ (these frame dependent fields are represented by the metric derivatives). The term
$dA_{i}/dt$ is a reactive force for noninertial motion and the term
$Curl(A_{i})=-2\Omega$ is the cosmic gravitomagnetic field $\vec{B}_{g}$, felt
only in frames rotating in the ever present matter-energy of the Universe at
the angular frequency $\Omega$ (figure \ref{fig:star-curl})~\cite{Unni-ISI}. This couples to moving masses,
angular momentum and spin (one should keep in mind that the gravitational
vector potential is really $-v\Phi_{u}/c$ and the induced relativistic field
is $\vec{B}_{g}=-2\Omega\Phi_{u}/c^{2}$, in the post-Newtonian language. A
spatially flat Universe, as observations confirm, implies that $\Phi_{u}%
/c^{2}=1$).

Thus \emph{we see that the electron's dynamics will be significantly affected
by the gravity of the matter-energy in the Universe, not included hitherto in
any consideration of physics}. In the specific case of a cyclotron orbit at
uniform speed, the term $\nabla\times\vec{A}_{g}=\vec{B}_{g}$ is the decisive
gravitomagnetic field generated by the dynamics in the gravitational presence
of the matter-energy in the Universe. \emph{Since the applied magnetic field
determines the cyclotron frequency }$\omega_{c},$\emph{ there is a relation
between the magnetic field and the cosmic gravitomagnetic field experienced by
the electron}.
\begin{equation}
\vec{B}_{g}=\nabla\times\vec{A}_{g}=-\nabla\times\vec{v}=-2\vec{\omega}%
_{c}=-2e\vec{B}/m
\end{equation}
Therefore, \textit{the cosmic gravitomagnetic field induced in every
electron's frame, due to the cyclotron orbital motion, is twice the cyclotron
frequency and it is directed opposite to the magnetic field}. This firm
result is the crucial physical input relevant for the QHE. Because of the
fixed relation to the cyclotron frequency, its coupling to the mass $mB_{g}$
can be replaced with $-2eB.$ Thus the mass of the electron drops out from the
effective expression, if written in terms of the applied magnetic field. This
is factually a coupling between the charge of gravity (mass) and a
gravitational potential, with a fictitious equivalence to the interaction
between the electric charge and a magnetic field, through the relation between
the cyclotron frequency and the applied magnetic field. In other words, the
induced cosmic gravitomagnetic field in electron orbitals is equivalent to
twice the (electro)magnetic field. \emph{This being a gravitational field, it
cannot affect or diminish the applied magnetic field}. Also, the induced
gravitational field vanishes if the electron's dynamics is arrested. However,
when it is present, it significantly affects the quantum phase and the
degeneracy of the LL, thereby completely changing the physical
phenomena in the LL.

Before going into the details of the new unified theory of the quantum Hall
effects based on cosmic gravity, we will now verify that the additional
quantum phase, due to the coupling of the relativistic gravitational potential
to the spin $\hbar/2$ of the electron, indeed solves the conundrum of the spin-phase
puzzle that I posed earlier. This will then prove the real existence of the
cosmic gravitational potentials.

\subsection{Proof of the reality of the cosmic gravitomagnetic field}

We can now \textit{re-examine the quantum phase in the spin-phase puzzle}.
Though I prove the reality of the cosmic gravitational effect through this
calculation, this does not figure explicitly in the gravitational theory of
the QHE because the electrons in the QHE situation are almost entirely spin
polarized. Since the working temperatures are below 0.1 K, and the magnetic
fields are well above 3 Tesla, the magnetic energy is 30-100 times the thermal
energy. With only the electromagnetic interaction of the electron magnetic
moment, we saw that the dynamical phase shift of the wavefunction in the
closed cyclotron orbital was $\delta\varphi_{s}=\pi,$ instead of the needed
$2\pi.$ Therefore, the condition on the phase closure in the orbit couldn't be
met in the cyclotron orbits of LL. \emph{The discrepancy is cured
exactly by the gravitomagnetic interaction}. The spin of the charge carrier in
the orbit couples to the induced cosmic gravitomagnetic field as $E_{s}%
=\vec{s}\cdot\vec{B}_{g}/2=-\vec{s}\cdot\vec{\Omega}$. The factor of 2
compared to the electromagnetic coupling $\mu\cdot B$ is due to the
different factors involved in the definition of the spin and the magnetic
moment. (The magnetic moment is $\mu=\frac{1}{2}\int\left(  r\times j\right)
dV$ whereas the spin or angular momentum is $l=\int r\times\bar{p}dV$. This
can also be understood in terms of the quadrupole (spin-2) nature of the
gravitational interaction). This gives a gravitational splitting $\pm
\hbar\Omega/2$. The lowest state is when the spin and the orbit are in the
same direction and the spin splitting is exactly equal to the cyclotron
splitting. This is important in comparing various theories (now onwards the
general angular frequency $\Omega$ is equated to the cyclotron frequency
$\omega_{c}$ and both symbols are used interchangeably).

With the gravitomagnetic field, we have the additional dynamical phase from
the energy of the spin of the electron in the gravitomagnetic field
$E_{g}=-\vec{s}\cdot\vec{B}_{g}/2$,%
\begin{equation}
\delta\varphi_{sg}=\frac{-\int\vec{s}\cdot\vec{B}_{g}dt}{2\hbar}=\frac
{-\int\vec{s}\cdot\vec{\Omega}dt}{\hbar}=-\frac{1}{\hbar}\frac{\hbar eB}%
{2m}\frac{2\pi}{\Omega}=-\pi
\end{equation}
This is really remarkable; \emph{the relativistically induced cosmic gravitational
potentials contribute exactly the phase required to ensure the quantum
condition on the phase}. The gravitational phase cancels the electromagnetic
phase and the phase closure condition of being an integer multiple of $2\pi$
is obeyed by the total quantum phase. Henceforth, we need not worry about the
phase associated with the spin. Note that the result of the correct closure
phase is independent of the relative sign of the spin and the magnetic moment;
if the two contributions happen to be of the same sign, the total phase is
still an integer times $2\pi$). \textit{This important result unambiguously
proves that the cosmic gravitomagnetic field is active in the physical system
of electrons in the LL. Thus, it is an essential physical input to formulate the
correct microscopic theory of the QHE.}

The relativistically induced cosmic gravitomagnetic field has important
consequences for several physical phenomena involving particles with a quantum
spin, when the dynamics is noninertial. The gravitational coupling to the
spin, with the energy $E_{s}=\vec{s}\cdot\vec{\Omega}\Phi/c^{2}$, can change
the quantum phase, split the energy levels, or cause a torque. Some of the
significant and definite results that I have already obtained pertain to the
spin-statistics connection, certain geometric phases of particles with a spin,
and the spin selectivity in the transport of electrons in chiral molecules
\cite{Unni-PhysNews,Naaman,Unni-NJP}.

The next important aspect is the magnitude of the induced cosmic
gravitomagnetic field in the cyclotron dynamics of every electron. It was
shown to be exactly $B_{g}=-2\Omega$, in the Universe at the critical density
of matter and energy, $\rho_{c}=3H_{0}^{2}/8\pi G$. This is equivalent to
setting $\left\vert \Phi_{u}\right\vert /c^{2}=1$. Since $\Omega=eB/m$ and the
gravitational potentials couple to the mass, the term $-2m\Omega=-2eB$. This
implies that the cosmic gravitomagnetic field, seen by the particle in the
Landau orbital enclosing $p$ units of the magnetic flux quanta, is equivalent
to exactly $2p$ units of the magnetic flux quanta, \emph{but directed opposite
to that of the magnetic field}. Those who have been working on the fractional
quantum Hall effect will no doubt recognize the importance of this result, in
the context of the association of the IQHE at $\nu=1$ and the FQHE at
$\nu=1/3$. \emph{We will see that the induced relativistic cosmic
gravitational field is behind the FQHE and its peculiar filling factors}.

While the induced field is equivalent to twice the magnetic field, one should
recognize that it is only an equivalence of the relative magnitudes, in the
specific context of the cyclotron motion. The relativistic gravitational field
induced in the electron's frame couples to the mass, and not to the electric
charge. \emph{Also, it cannot diminish or cancel the applied magnetic field in
any measure}. The summary statement is that the induced gravitational field
affects the quantum phase of the electron twice as efficiently as the magnetic
field in this situation. However, it is clear that there is no gravitational
force; only the dynamical action and the phase of the wavefunction are
modified due to the interaction. I stress that this \emph{cosmic
gravitomagnetic field (denoted gravitomagnetic) is as real as the
matter-energy in the Universe and cannot be ignored or denied, because the
gravitating matter in the Universe is factually observed and measured}.

Note that in this derivation, we have used the well tested equality of the
gravitational mass and the inertial mass; otherwise the ratio $m_{g}/m_{i}$
will multiply the gravitomagnetic field $\Omega$ (this equality is derived in
the theory of Cosmic Relativity \cite{Unni-WAG}). The concept of a flux
quantum is paraphrasing the quantization of the closure phase of the
wavefunction in the presence of a field. It is perhaps superfluous to
emphasize that the flux of $\Omega$ or of the magnetic field $B$ is not intrinsically quantized; it
is the quantum phase of the electron's dynamics that is quantized.

This shows why the theories invoking composite particles had to postulate a
fictitious gauge field (unphysical, meaning not of any fundamental
interactions) related to the multi-particle effects, to arrive at the correct
phenomenology. \emph{In the Composite Fermion theory of the FQHE, the
postulated field dresses the carriers with an even number of `inverted' flux
quanta and makes complex composite particles that diminish the real magnetic
field}. In reality, there are no composite particles involved in the
phenomenon. Neither is there any cancellation of the magnetic field. There are
two long range fields -- electromagnetic and gravitational -- acting
simultaneously. One couples to the electric charge and the other to the mass.
This is the crucial point; the quantum phase of the particles changes due to
the magnetic potential coupling to the electric charge and the gravitational
potential coupling to the mass. The flux quantum of the magnetic field is
$h/e$ and that of the gravitomagnetic field is $h/m_{g}$. The response to both
the interactions, however, is determined by the mass (inertia). Therefore, one
term has the coefficient $e/m$ and the gravitational term has the coefficient
unity, $m_{g}/m_{i}=1$. \emph{The net effect on each particle can be written
formally in terms of the magnetic field because the effect of the
gravitomagnetic field is through }$\Omega$\emph{, the cyclotron resonance
frequency, which is determined by the magnetic field and the charge-to-mass
ratio}. Hence, all the elements of the microscopic physics in the situation
are at hand now. The gravitational scenario of the quantum Hall effects (GQHE)
does not have any extraneous hypothesis or free parameter. 

\subsection{Calculation of the degeneracy of the Landau levels}

What determines the degeneracy of the quantum levels? Conventionally, the
degeneracy of a single Landau level in the 2D system is calculated in a
specific choice of the gauge of the magnetic vector potential. This affects
the geometry of the orbitals, but the degeneracy $G_{q}$ is of course
independent of the gauge. Here, I will use a direct method that relies on the
core physical reason that determines the quantum degeneracy, which is the
`action' $S$ associated with the electromagnetic interaction that determines
the physical orbitals \cite{Unni-RQM}. This is very similar in spirit to the method used by S.
N. Bose in the calculation of the degeneracy in the derivation of the Planck
radiation formula \cite{Bose-1924}. The simplicity of the derivation speaks
for its fundamental robustness. The quantum degeneracy is simply the available
`action space' that is quantized in the units of $h.$ Then the total quantum
degeneracy is $G_{q}=S/h$ in all physical situations. The electromagnetic part
of the action for the LL in the magnetic field $B$ is $\int eAdl$
and the degeneracy density (per unit area) is%

\begin{equation}
g_{m}==\frac{1}{h}\int eAdl=\frac{1}{h}\int\int e\left(\nabla\times
A\right)  \cdot d\Sigma=\frac{eB}{h}=\frac{B}{\phi_{0}}%
\end{equation}
I have shown that the degeneracy density of the LL is universally
the density of the flux quanta. This particular derivation also shows an
important physical fact that has relevance to the experiments in which the
field is applied at an angle $\theta$ to the normal. Then the degeneracy
density is reduced by the factor $\cos\theta$. However, the degeneracy is
obviously not sensitive to the sign of the field.

\emph{I now show the most important result that is relevant for the new
physical theory of the quantum Hall effects, that the degeneracy in the LL
is modified by the relativistic cosmic gravitational potential in the
electron's orbital frame}. With the cosmic gravitational potential
$g_{0i}=-v_{i}/c$, the change in the gravitational part of the action for
every electron present is $S_{g}=-\int mv_{i}dl$. The degeneracy due to the
applied magnetic field does not depend on the number of carriers, but the
gravitational part is proportional to the number of the electrons that are
factually present in the Landau orbitals, because the gravitational
potentials are relativistically induced by the dynamics of the electrons. Then
the new degeneracy density per level is the average over the ensemble of electrons in
the level. The degeneracy per unit area in a LL is then
\begin{align}
g &  =\frac{eB}{h}-\frac{2\nu^\prime}{h}\int\int m\Omega\cdot d\Sigma\nonumber\\
&  =\frac{B}{\phi_{0}}-2m\nu\Omega/h=\frac{B}{\phi_{0}}-2\nu\frac{B}{\phi_{0}%
}=\frac{B}{\phi_{0}}-2\rho^\prime
\end{align}
Here, $\nu^\prime$ is filling factor in a particular level (between 0 and 1) and $\rho^\prime$ is the corresponding number density in that level. The maximum gravitational contribution to the degeneracy is $2m\Omega/h$ (when $\nu=1$), exactly analogous to the magnetic contribution $eB/h$. When the carrier density is zero, there is no contribution, signifying its dynamical origin. 
\begin{figure}
	\centering
	\includegraphics[width=0.8\linewidth]{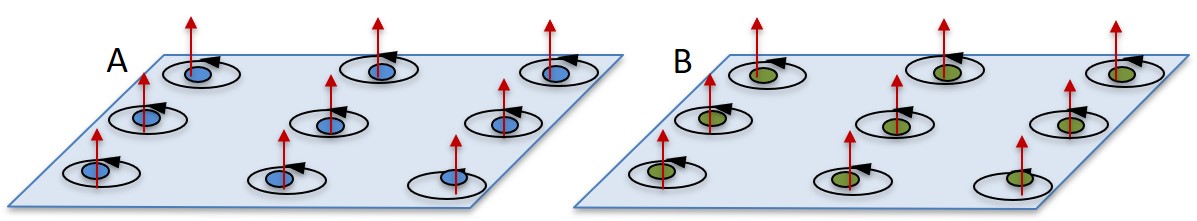}
	\caption{The degeneracy at the complete filling of a LL is not altered by the gravitational effect. A) The degeneracy $g_m$ with only the magnetic field considered, $B/\phi_{0}\equiv b$. At the complete filling, $\rho=g_m$ one has the integer QHE. The arrows represent the magnetic field. The circles represent the electrons in dynamics. The filled blue blobs represent the degeneracy. The figure is schematic and does not depict the spatial situation in the material. B) The induced gravitomagnetic field can alter the degeneracy as $g=g_m-2\rho$. Thus, the average degeneracy is really $b-2\rho$. But, at the complete filling of the levels, $\rho=b$ and $g=\left\vert b-2b\right\vert =b$. The degeneracy remains the same in spite of the modification of the dynamics (quantum phase) due to the induced gravitomagnetic field.}
	\label{fig:iqhe}
\end{figure}
\emph{The quantum degeneracy depends now on
both the applied field} $B$ \emph{and the carrier density $\rho^\prime$ in the level}. Three
remarkable facts about the QHE are evident in this discovery right away.
\begin{enumerate}
	\item When an energy level is completely filled, $\nu=1$, the degeneracy remains
	unchanged in the induced gravitomagnetic field; it remains at $\left\vert
	B/\phi_{0}\right\vert $, showing that the gravitational interaction that is
	present does not alter the integer quantum Hall effect. As long as the level
	remains filled, this situation continues (figure \ref{fig:iqhe}).
	\item When there is
	exactly half-filling, $\nu=1/2$, the field-related degeneracy is zero and the
	only possibility is a Fermi sea of the electrons. 
	\item \emph{For several values of the fractional filling, the carrier density can be an integer multiple of the effective degeneracy}. Then the QHE of the electrons occurs at the
	fractional filling factors, exactly as in the integer effect. Since both
	$\nu=p/q$ and $\nu=1-p/q$ result in the same degeneracy, the QHE will occur at
	both these fractions when $\rho/g$ is an integer. But, the QHE manifests
	differently at these filling factors because the factor $\rho/g$ is different
	in the two cases. 
\end{enumerate} 
\emph{These results already explain the main patterns of the
entire terrain of the quantum Hall effects, within a unified single particle
quantum theory. }

I emphasize the physical fact that the factual degeneracy is modified because
of the induced gravitomagnetic field in the cyclotron orbits. The source of
the relativistic gravitational field is the motion relative to the cosmic
matter that verifiably exists, and is already observed. So far we have not
introduced even a single extraneous hypothesis, and we will continue to
explore the immediate consequences for the QHE.

\section{The Integrated Theory of the Quantum Hall Effects}
I emphasize the results by summarizing the gravitational scenario. The
relativistically induced cosmic gravitomagnetic potential modifies the
dynamical action and the degeneracy of the quantum energy levels, making the
degeneracy a function of both the applied magnetic field and the factual
carrier density in the two dimensional material. This is the real cause of the
QHE at the fractional filling factors. The phenomenon at the prominent filling
factors for $\nu\geq1/3$ is exactly the same as in the integer quantum Hall
effect; there are no significant multiparticle effects or any quasiparticle
involved. Nor is there any internal modification of the applied magnetic
field. All significant experimentally observed peculiarities must be
explainable by the gravitationally modified degeneracy. For $\nu<1$, the
available energy levels are of course determined by the Coulomb interaction.

The magnitude of the degeneracy, in terms of the LL filling factor $\nu^\prime
=\rho^\prime/\left(  B/\phi_{0}\right)  $ and the flux quantum density $b=B/\phi_{0}%
$, is%

\begin{equation}
g=\left\vert \frac{B}{\phi_{0}}-2\rho^\prime\right\vert =\left\vert b\left(
1-2\nu^\prime\right)  \right\vert \label{degeneracy}%
\end{equation}
When $\nu^\prime=1$, we see that the degeneracy per unit area, $g=\left\vert
-B/\phi_{0}\right\vert \equiv b$, is the same as when only the magnetic field
is considered for the degeneracy (figure \ref{fig:iqhe}). This is because the
induced gravitomagnetic field is in the direction opposite to the applied
magnetic field for the electron. Since the effective gravitational action is
exactly twice the action from the magnetic interaction, but with the opposite
sign, \emph{the magnitude of the degeneracy is unaltered}. This is the
IQHE with $\nu=1$. When the number density of the particles is larger, they
occupy the next higher Landau energy level, in the same magnetic field. At $\nu=2$,
the second level also has the number density equal to the degeneracy, and the
IQHE at $\nu=2$ follows. Thus, all the filling fractions at which the IQHE
effect happens are fully explained without any modification of the theory.
Next, we examine the situation when $\nu<1$ (Then, there is only the lowest LL to consider; $v^\prime=\nu$ and $\rho^\prime=\rho$).

\subsection{The microscopic equivalence of the IQHE and the FQHE}
\begin{figure}
	\centering
	\includegraphics[width=0.8\linewidth]{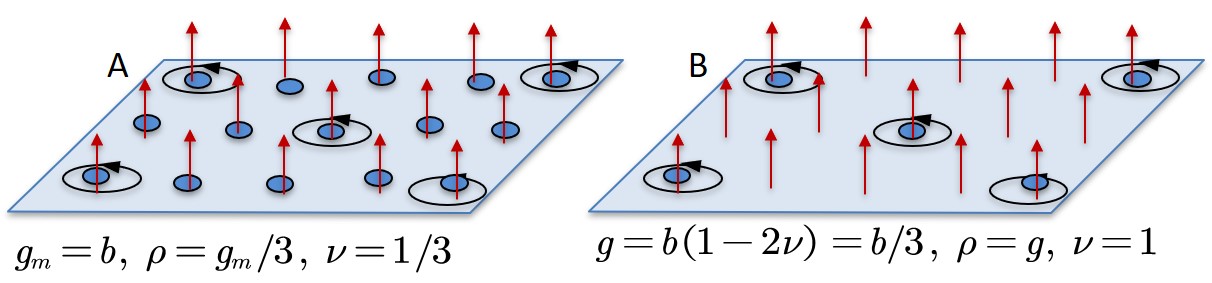}
	\caption{The gravitational action that realizes the fractional quantum Hall effect at the filling fraction $\nu=1/3$.  A) The conventional view of the fractional filling of the degeneracy, one electron per 3 flux quanta ($b=B/\phi_{0}$) on the average. The arrows represent the magnetic field and the blue blobs represent the degeneracy. The circle represents the electron. The degeneracy in this case is equal to the flux quantum density, $g_m=b$. B) The induced gravitational field alters the degeneracy, $g=b(1-2\nu)$. For $\nu=1/3$, the quantum degeneracy is modified to $g=b/3$. The magnetic field is not altered. Then the number density of electrons exactly fills the degeneracy, resulting in the simple integer QHE.  This is the FQHE at $\nu=1/3$. There are no fractionally charged quasiparticles or particle-flux composites involved in the phenomenon.}
	\label{fig:fqhe}
\end{figure}

In the single particle theory, the IQHE can occur only if the degeneracy $g$
at some magnetic field is \emph{smaller} than the carrier density $\rho$ such
that the ratio $\rho/g$ is an integer. Without the gravitational effect, this
possibility does not exist because $g=g_{m}=b$ and $\rho/b=\nu<1$. But, the
situation has drastically changed with the gravitational effect. We have
\begin{equation}
\rho/g\equiv\nu^{\ast}=\frac{\rho}{b\left(  1-2\nu\right)  }=\frac{\nu
}{\left(  1-2\nu\right)  } \label{qhe-ratio}%
\end{equation}
I emphasize again that this expression was obtained from the factual
relativistic physics, with the magnetic field coming from the laboratory
currents and the gravitational effect coming from the motion relative to the
observationally confirmed matter-energy in the Universe. There is no new
assumption or new hypothesis involved. \emph{There is no modification of the
magnetic field felt by the charged particle, and there is no modification of
the effective charge}. In other words, we are dealing with independent
electrons and their expected Coulomb interaction with the energy $E_{C}\approx
e^{2}/\epsilon l$, and there are no quasiparticles constructed from the
multi-particle effects. What is modified by the extra gravitational potential
and the corresponding action is the degeneracy of the states in the LL. The degeneracy has become dependent on the filling fraction in the
level -- instead of $b=B/\phi_{0}$ it has become $b\left(  1-2\nu\right)  $.
This simple physical fact has the remarkable consequence that the QHE occurs
at fractional filling factors (figure \ref{fig:fqhe}).

\begin{figure}
	\centering
	\includegraphics[width=0.8\linewidth]{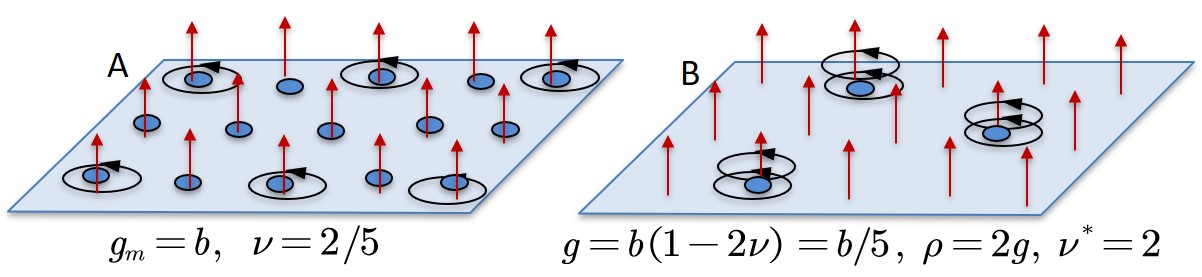}
	\caption{The illustration of the fractional quantum Hall effect at the filling factor $\nu=\rho/\left( B/\phi_{0}\right) =2/5$. A) The conventional view of the fractional filling, 2 electrons for 5 flux quanta. There cannot be the QHE at this partial filling. B) The induced gravitational field drastically reduces the degeneracy to $g=b(1-2\nu)=b/5$. The magnetic field is not altered. Then there are two electrons per degeneracy slot that are distributed in two energy levels defined by the Coulomb interaction in the material. This is the FQHE at $\nu=2/5$.}
	\label{fig:qhe-25}
\end{figure}

We see that $\nu^{\ast}$ is an integer $n$ for all the fractions $p/q$ that
satisfy
\begin{equation}
	\pm n=\frac{p/q}{1-2p/q}\Longrightarrow\frac{p}{q}=\frac{n}{2n\pm1}%
\end{equation}
This predicts the primary fractions for $\nu<1$ as $p/q=2/3,3/5,4/7...$ and
$p/q=1/3,2/5,3/7...$ (figures \ref{fig:fqhe},\ref{fig:qhe-25}). Of course, \emph{this also predicts the fractional QHE
	for} $\nu>1,$\emph{at all the fractions }$i+p/q$\emph{, like }%
$5/3,8/5,...4/3,7/5...$\emph{ etc. and many more with} $i=1,2,3...$.  This is
because when a particular Landau energy level is already completely filled, it
remains filled until the higher level is completely depleted; the change in
the degeneracy with the magnetic field is compensated by the filling from the
upper level.

\begin{figure}
	\centering
	\includegraphics[width=0.8\linewidth]{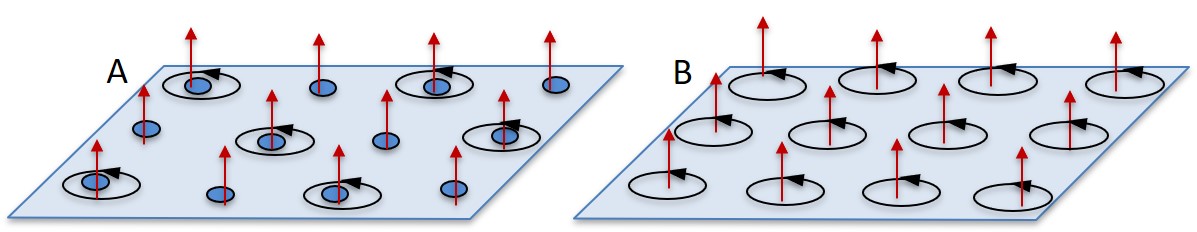}
	\caption{The real physical situation at the half-filling of a LL. A) With only the magnetic field (arrows) considered, the degeneracy density (blue blobs) is $g=B/\phi_{0}$, and the carrier density is $\rho=g/2$. B) The cosmic gravitational effect renders a very different situation: the magnetic degeneracy is modified as $g^{\prime}=\left\vert 2\rho-B/\phi_{0}\right\vert =0$ at which the QHE is impossible. The magnetic field experienced by the electrons remains as $B$. Only the degeneracy is modified, and not the magnetic field.}
	\label{fig:half-fill}
\end{figure}

We saw that the gravitational contribution to the degeneracy (eq.
\ref{degeneracy}) cancels the magnetic part exactly at $\nu=1/2$ resulting in
the impossibility of any QHE near that filling factor. \emph{It is important to note
also that the longitudinal magnetoresistance in this theory near $\nu=1/2$ is
what one expects from the classical Hall effect, at the actual applied
magnetic field $B$}. This is because the magnetic field sensed by the charged
particles is the full applied field without any modification. The
gravitational effect only serves to modify the action and the degeneracy of
the levels (figure \ref{fig:half-fill}). This unique prediction agrees well with what is
observed in measurements. Thus, the major features of the fractional QHE up to
the filling fraction $\nu=1/2$ are straightforward and natural in this simple
single particle description of the quantum Hall effect (we will discuss the
minor fractions in this range soon). Going further, we see that the QHE
happens prominently also beyond $\nu=1/2$, at the filling fractions in the
sequence $4/9,3/7,2/5$, reaching at the `golden fraction' $1/3$. When
$\nu=1/3$,  the degeneracy matches the particle density again, $\rho
/g=\nu^{\ast}=\nu/(1-2\nu)=1$, similar to the case of the IQHE at $\nu=1$. 

At this stage, the enormous simplicity of the gravitational paradigm and the
stark contrast with the previous attempts for a microscopic theory of the QHE
should be noticed. The unified single particle theory with the necessary
gravitational contribution explains the entire sequence of the integer filling
fractions and all the prominent fractional filling fractions up to $\nu=1/3$
at which the QHE happens. The theory also explains the absence of the QHE at
the filling fractions $\nu=1/2$, and further at $\nu=1/4$. \emph{It is clear
that the FQHE is exactly the same physical phenomenon as the IQHE}. There is
no difference in the microscopic physical description. There are no new
quasiparticles or a new gauge field involved in the QHE at the fractional
filling factors. The entire sequence of the QHE at the integer and fractional
filling factors results from the single fact of the dual dependence of the
quantum degeneracy, on the applied magnetic field as well as on the
relativistically induced gravitomagnetic field. The gravitational contribution
is directly related to the actual number density of the particles in the
cyclotron dynamics.

For $\nu<1$, the QHE sublevels in the lowest LL implied by the
fractions like $\nu=2/3,\nu^{\ast}=2$, $\nu=2/5,\nu^{\ast}=2$, $\nu
=3/7,\nu^{\ast}=3$, require explanation. Since we are already in the lowest
LL, the multiple energy levels that are occupied must be the quantum
states separated by the Coulomb interaction between the electrons, with the
mean energy given by the energy at the mean separation at the actual spatial
filling. Thus, the multiple energy levels in the lowest LL in the
GQHE theory have no relation to any modified cyclotron frequency $\Omega
^{\ast}$, unlike in the CF theory that relies on a diminished magnetic field
and the corresponding energy levels separated by $\hbar\Omega^{\ast}$. We will
examine this later, in the context of the very important thermal activation
measurements of the energy gaps.

We immediately note that the smaller fractions like $1/5,2/7$ and $1/7$ are
missing from the list that is derived directly from the gravitational
scenario. Given the fact that the system has a gravitational QHE state at
$\nu=2/5$ which corresponds to $\nu^{\ast}=2$, one might expect that exactly
the same configuration with $\nu^{\ast}=1$ is the $\nu=1/5$ FQHE state.
However, I want to state an important caveat here, because of which a deeper
analysis of such fractions is required within the GQHE theory. There is no
natural way, without any additional hypothesis, to get fractions like $1/5$ or
$1/7$ in the gravitational scenario, only using the first principles.
Similarly, fractions like $4/5$ or $5/7$ are not natural, as primary
fractions. Yet, we know that some of such fractions have been observed and
they need to be explained adequately without arbitrary postulates. We will
discuss at the end the minimal gravitational scenario that might explain such fractions.

\subsection{The 5/2-FQHE state and the half-integer states}

Without the mention of the enigmatic `5/2-FQHE state', one cannot complete the
discussion of the fractional QHE for the filling factors above $\nu=1$
\cite{Willett-FiveTwo} There is no conclusive theoretical picture till now for
the 5/2-FQHE state (spin polarized), which happens at the second Landau level,
$\nu=5/2=2+1/2$. This state does not fit in the gravitational physical scheme
either, in a natural way. In fact, \emph{the QHE should not happen at any
half-filling, in the gravitational paradigm}, if the carriers are
spin-polarized. The 5/2-FQHE state is seen only in some samples under special
experimental situations. The magnetoresistance behaviour at $\nu=7/2,9/2...$
etc., is generally similar to that at $\nu=5/2$. A rare 3/2 state is also
studied. 

\begin{figure}
	\centering
	\includegraphics[width=0.7\linewidth]{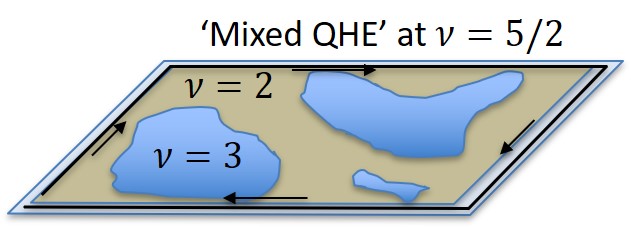}
	\caption{A schematic representation of the mixed quantum Hall state at $\nu=5/2$, arising as a spatial average of the legitimate QHE states that straddle $\nu=5/2$. The even fraction results from the spatial averaging. Here, comparable contributions of $\nu=2$ and $\nu=3$ are shown. The longitudinal resistivity vanishes for both states.}
	\label{fig:mixed}
\end{figure}

The simplest single particle picture then is that the physical behaviour at
$\nu=5/2$ is originating from a mixed state of $\nu=(2,3)$, due to the local
microscopic fluctuations and the spatial mixing between the $\nu=2$ and the
$\nu=3$ IQHE states.\emph{ In other words, there is no 5/2-FQHE}. Different
spatial regions of the sample are locally stabilized in the $\nu=2$ and
$\nu=3$ filled levels (figure \ref{fig:mixed}). This prompts us to claim that the
$\nu=5/2$ state is really the $\nu=(2,3)$ mixed state and not a new kind of
fractional QHE state. This need not be directly a mixed state $\nu=(2,3)$; it
can also be the mixture of more nearby $\nu=7/3=2+1/3$ and $\nu=8/3=2+2/3$.
What seems to be seen in the experiments at ultra-low temperatures in very
clean samples is the mixing of the IQHE states $\nu=2$ and $\nu=3$ or of the
states $\nu=7/3$ and $\nu=8/3$, at the intermediate field. Since the measured
resistivity is a spatial average over the sample, one gets an intermediate
flat plateau, as well as the vanishing longitudinal resistivity. Only detailed
experiments can settle the issue. One experimental possibility is to cycle the
field between $\nu=2$ and $\nu=3$, and look for the rearrangement and the
annealing of the spatial mixing. We expect the $\nu=3/2$ state to be more
difficult to realize because of the wider gap, in the magnitude of the
magnetic field, between the $\nu=1$ and $\nu=2$ states. But, since the states
$\nu=5/2$ and $\nu=3/2$ are similar, both are possible if the spatially mixed
states of nearby filling factors indeed occur in the integer QHE.

\section{The Challenge Beyond $\nu=1/3$}

The gravitational single particle theory with the cosmic gravitational effect
on the quantum degeneracy explains the filling factors at which the QHE
occurs, for $\nu\geq1/3$. The QHE has been observed also for the filling
fractions less than $\nu=1/3$ and it is obvious that the gravitational theory
has to consider some new aspects to explain these fractions. The sequence
beyond $\nu=1/3$ is $2/7,3/11,4/15,...4/17,3/13,2/9$ etc. Of course, phenomenologically
the entire sequence all the way to $\nu=1/5$ is reproduced by
\begin{equation}
\rho/g=\frac{\nu}{\left(  1-4\nu\right)  }%
\end{equation}
where the modification of the degeneracy is $b\left(1-4\nu\right)$ instead
of $b\left(1-2\nu\right)$. This corresponds to a doubling of the quantum
phase due to the gravitational interaction. Normally, the dynamically induced gravitomagnetic
field $2\Omega$ corresponds to $-4\pi$ in the quantum phase. Combined with the
$2\pi$ phase of the magnetic interaction, this results in the total phase of
$-2\pi$, which is equivalent to the original $2\pi$ with the magnetic field
alone. This is how the integer QHE is preserved as it is in the gravitational
paradigm, and the fractional QHE is facilitated when $1/3<\nu<1$. For
$1/5<\nu<1/3$, the dynamics of the electron should be such that the gravitational contribution modifies the quantum degeneracy
to
\begin{equation}
g=\left\vert \frac{B}{\phi_{0}}-4\nu\frac{B}{\phi_{0}}\right\vert =\left\vert
b\left(  1-4\nu\right)  \right\vert
\end{equation}
Then one can get the entire sequence of the FQHE plateaux, $2/7,3/11...$ till
$\nu=1/5$.

However, there seems no obvious reason why the contribution to the action
corresponding to the induced gravitomagnetic field is twice its natural value
of $-2\nu\Omega$ per unit area for the filling fractions $\nu<1/3$, up to
$\nu=1/5$. This is the first instance when I am tempted to consider an
extraneous hypothesis in the gravitational paradigm for the QHE. But, in this paper, I
refrain from perturbing the remarkable edifice and robust structure of the
gravitational paradigm of the QHE. My conviction is that the reason for this
will emerge eventually, because the cosmic matter and its gravity are
undeniable and the gravitational effects discussed for $\nu\geq1/3$ are the
natural consequences. These remarks apply also to the fractions further down
in the sequence, like $\nu=1/7$. Instead of forcing the resolution
with an opportunistic hypothesis, the correct physical reason needs to be
found, in relativistic gravity and nonrelativistic quantum mechanics.

\section{The Spin and the QHE}

At the magnetic fields and temperatures at which the QHE is typically observed
(several Tesla field and temperature below 100 mK), the sample is essentially
spin polarized in the lowest state. The Zeeman energy gap is about 10 times
the thermal energy, even in the lower range of magnetic fields. In the GQHE
scenario, the magnetic field is not modified (diminished) by the
gravitomagnetic field, obviously. Hence, the cyclotron frequency is not modified.

The new important physical factor is the gravitomagnetic energy, due to the
interaction of the spin with the cosmic gravitomagnetic field, $E=s\cdot
B_{g}/2=s\cdot\Omega$. The factor of 2 compared to the electromagnetic
interaction $E=\mu\cdot B$ is due to the difference in the definition of the
magnetic moment and gravitomagnetic moment (spin). Just as the mass is the
`charge' (source) of gravity, \emph{the spin in gravitational physics is the
equivalent of the magnetic moment in electromagnetism}. The difference in
the \emph{gravitational energy} of the two spin states is thus $\Delta
E=\hbar\Omega$, with $\Omega=eB/m^{\ast}$. This equals the cyclotron gap and
far exceeds the Zeeman splitting $\Delta sg^{\ast}\mu_{B}B$. We already saw
how this interaction and the resultant dynamical phase are essential for
fulfilling the correct quantization condition in the Landau orbitals. They are
also important in determining the filling sequence in the Landau levels.

Though not manifest in the QHE at high magnetic fields, the interaction of the
quantum spin with the induced cosmic gravitomagnetic field is the physical
reason for many phenomena, considered as intriguing hitherto. As I mentioned
earlier, certain geometric phases measured with polarized photons and neutrons
\cite{Tomita-Chiao,Neutron-geom-phase}, as well as the puzzling spin
selectivity seen in the transport of electrons in chiral molecules
\cite{Naaman}, are important examples \cite{Unni-NJP}. In fact, the induced cosmic gravitomagnetic field experienced by the electrons transported in chiral molecules is enormous, equivalent to more than a thousand Tesla. This results in a strong spin selectivity in the conduction, even at the room temperature, due to the large gravitational energy gap $E=s\cdot\Omega$ between the two spin states. A significant advance in fundamental physics is the finding that the physical reason for the elusive
spin-statistics connection is the interference of the relevant amplitudes with the gravitational quantum phase, $\int Edt=2\int s\cdot \Omega dt$, due to the interaction of the
spin with the induced cosmic gravitomagnetic field \cite{Unni-spinstat,Unni-PhysNews}.

\section{The Excitation Energy Gaps}

\begin{figure}
	\centering
	\includegraphics[width=0.8\linewidth]{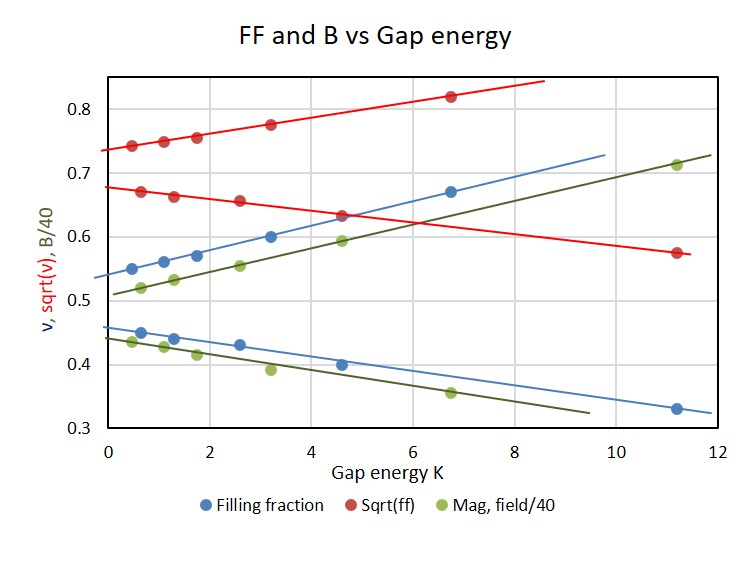}
	\caption{The approximate linear correlations between the measured energy gaps $\Delta_{\nu}$ and the magnetic field ($B$, green), the filling factor ($\nu$, blue), and the square root of the filling factor (red). (The $B$-values are scaled by 40, to plot in the same scale). The two lines for each parameter factually straddle the point where $\nu=1/2$, though here they are shifted and scaled, for accommodating in a single plot with clarity. The general belief was that the linear correlation with $B$ strongly supports the CF theory.}
	\label{fig:gaps}
\end{figure}

The energy gaps in the lowest LL at the various filling fractions
have been measured from the temperature dependence of the longitudinal
resistivity \cite{Stormer-gaps}, assuming
\begin{equation}
\rho_{xx}=\rho_{0}e^{-\Delta/2kT}%
\end{equation}
The data on the energy gaps at several  filling factors in the FQHE regime follow an approximately linear dependence on the `effective magnetic field' $\Delta B_{\nu}=B_{\nu}-B_{1/2}$, taking the field at $\nu=1/2$ as the zero point. This lends good support to the CF theory in which the effective magnetic field felt by the carriers is a modified magnetic field that vanished at $\nu=1/2$. I consider that these measurements are very important for building an accurate
theory. However, the linear dependence of the measured gap
energies on $\Delta B_{\nu}$  is misleading for the following reason.

A similar approximate linear dependence, with the same qualitative goodness,
may be verified if the gap energies are plotted against the filling factors
$\nu$ themselves, or even against the square root of the filling factors
(figure \ref{fig:gaps}). This just means that the range of the measurements is
not enough to draw an unambiguous conclusion. Evidently more measurements in
this genre are required to formulate a reliable picture of the energy gaps and
activated transport in the lowest LL. The magnitude of the gaps is
certainly consistent with the estimates of the Coulomb energy at the carrier
density and the filling factors. Yet, there are multiple possibilities for
microscopic physical configurations. The magnetic length $l_{0}$ is related to
the inverse of the degeneracy in the magnetic field $B$, we have $l_{0}%
=\sqrt{2\pi\phi_{0}/B}=\sqrt{2\pi/g_{m}}$. Then the Coulomb energy is
$E_{C}\approx e^{2}/\epsilon l_{0}\propto e^{2}\sqrt{g_{m}}$. Since the FQHE
occurs when the filling fraction is an integer times the degeneracy, $\nu=ng$,
we get
\begin{equation}
E_{C}\propto\frac{1}{\sqrt{n}}e^{2}\sqrt{\nu}
\end{equation}
It is clear that speculating with the limited data will not be conclusive.
However, what is learned from this exercise is that the data on the activation
gaps are consistent with the gravitational paradigm of the quantum Hall effects.

\section{Electron-Hole Symmetry and the FQHE}

Similar to the classical Hall effect, the QHE is observable with both electron
and hole carriers. This is most dramatic in monolayer graphene, where both
branches appear symmetrically as the gate voltage (carrier density) is scanned
on both sides of zero. The hole is already a `quasiparticle' -- an equivalent
single positively charged particle of a complex multiparticle transport of
electrons in the valence band. Though the spatial environment of both
electrons and holes is similar -- the crystal lattice with many electrons --
their interaction environments are very different.

If the carriers are holes, the orbits in the magnetic field have the opposite
sense compared to electrons. (In reality, it is the electrons that have the
dynamics, but the total dynamics is shared by many electrons, each
contributing as single particles). The gravitomagnetic field is determined by
the sense of the orbital motion and that also reverses sign, making the
gravitomagnetic field induced by the dynamics of the `hole' notionally
oriented in the same sense as the magnetic field. However, the magnetic
degeneracy is now with a positive charge and therefore, the sign of the
magnetic flux seen by the `hole' quasiparticles remains in the opposite sense
to the gravitomagnetic field. So, the GQHE scenario carries over consistently
to the Hall effect for the holes as well, without any additional hypothesis.

In the present phenomenological theories of the FQHE, the situations involving
the hole conductivity need additional assumptions and statements, as expected
in the `epicyclic' and hierarchical theories. For example, in the Composite
Fermion picture, the notion of the flux quanta attaching to the hole-charge is
a concept `twice removed' in that both the `hole' charge and the flux quanta
are fictitious. This issue with the difficulties in a microscopic physical
picture was indicated earlier.

\section{Quantum Hall Effects in Graphene}

A single atomic layer of graphene is an ideal 2D material \cite{Graphene-rev}
to explore the microscopic aspects of the theories of the QHE. Unlike in the
semiconductor heterostructures that have the 2D `gas of electrons', the
transport and conduction of the electrons in graphene are dominated by the
nearest neighbour hopping. Also, owing to the peculiar `light-come' dispersion
structure, the energy levels in the presence of the strong magnetic field are
quantized very differently compared to the levels in the other 2D materials;
the quantized energy levels (for electrons and holes) are
\begin{equation}
E_{n}=\pm\sqrt{2}\hbar v_{F}\sqrt{n}/l_{0}=\pm\hbar v_{F}\sqrt{\frac
{2eBn}{\hbar}}=\pm v_{F}\sqrt{2\hbar eBn}%
\end{equation}
instead of the LL quantization $E_{n}=\hbar eB\left(  n+1/2\right)
/m$ for the 2D electron gas (2DEG). Yet, the integer QHE in graphene is very
similar to the QHE in other 2D materials. This is not surprising in the
gravitational paradigm of the QHE. In fact, the physics of energy quantization
is not very different from the LL quantization. Usually, the
quantization of energy levels in graphene is discussed in the context of the
relativistic dynamics of a massless particle and the Dirac equation. However,
the quantum orbitals of the nonrelativistic dynamics in the magnetic field do
not correspond to free motion; the orbitals are realized by the site to site
hopping of the electrons in the graphene lattice. \emph{The mean velocity is
determined and fixed as a constant by the hopping frequency} $\nu_{h}$ and the
lattice spacing $a$, as $v_{F}\approx a\nu_{h}$. The quantization of the
energy can be determined directly from the quantization of the action
\cite{Unni-RQM}. Since $\int E_{n}dt/\hbar$ is an integer multiple of $2\pi$,
we have%
\begin{equation}
2\pi n=\frac{1}{\hbar}\int E_{n}dt=\frac{2\pi E_{n}}{\hbar\omega_{c}}%
=\frac{2\pi E_{n}r_{n}}{\sqrt{2}\hbar v_{F}}%
\end{equation}
Here, $r_{n}$ is the notional radius of the cyclotron orbital. I have used the
relation that is valid in graphene, $\omega_{c}=\sqrt{2}v_{F}/r_{0}$, which
has the extra $\sqrt{2}$ factor compared to the expression for free cyclotron
orbits. Now, the increments in the radial size of the closed orbitals should
obey the condition that the increment in the area corresponds exactly to one
flux quantum. Therefore, $r_{n}^{2}=nr_{0}^{2}$, where $r_{0}$ is the
`magnetic length'. The energy levels are%
\begin{equation}
E_{n}=\frac{\sqrt{2}\hbar v_{F}n}{r_{n}}=\frac{\sqrt{2}\hbar v_{F}n}{\sqrt
{n}r_{0}}=\hbar v_{F}\sqrt{2eB/\hbar}\sqrt{n}=v_{F}\sqrt{2\hbar eBn}%
\end{equation}
\emph{Therefore, the }$\sqrt{n}$\emph{ quantization in graphene is a
consequence of the fixed velocity }$v_{F}$. In the 2DEG, the velocity changes
with the cyclotron radius and energy since the cyclotron frequency is a
constant, $\Omega=v/r$.

Since it is very easy to change the filling fraction $\nu=e\rho/hB$ by
changing the number density of carriers, which is linear in the applied gate
voltage, the QHE in graphene is very often observed in a constant magnetic
field. Of course, the quantum degeneracy due to the applied magnetic field is
exactly as in the 2DEG, $g=B/\phi_{0}$. Since the $n=0$ level is of zero
effective carrier density, with the Dirac point equally shared by the
electrons and holes, there is no plateau at $n=0$. Each level has the further
2-fold degeneracy for the two spin projections and another 2-fold degeneracy
from the two sublattices in the graphene structure. The relation for the
complete filling of the levels is then $\rho/\left(  B/\phi_{0}\right)
=4\left(  \nu+1/2\right)  $, with $\nu=0,1,2...$ etc.; the Hall conductivity
is quantized in units of $4e^{2}/h$.

Already the integer effect in graphene raises some unique questions about the
theory of the QHE. One issue is the essential role of the impurities for
localizing the electronic states, which is vital for the theory of the Hall
plateaux. The concentration of the impurities and their distribution are
certainly very different in the single atomic layer graphene, and yet, the QHE
happens in a robust manner, even at a relatively high temperature. Also, the
behaviour of the longitudinal conductivity $\sigma_{xx}$, as the carrier
density is varied, invites more studies on the interplay between the localized
states and the extended states in the Landau levels.

The observation of the fractional QHE in monolayer graphene allows the close
scrutiny of the theories of the FQHE based on the hypothetical quasiparticles.
Many QHE fractions have been observed in the measurements in high magnetic
fields, where the 4-fold degeneracy is lifted. The prominent fractions
$1/3,2/3,3/4,4/7...$ etc. between $\nu=1$ and $\nu=1/3$, as well as several
fractions like $4/3,5/3,7/3...$ etc. for $\nu$ between 1 and 3 have been
observed. Also, the energy gaps at some fractions have been deduced from the
measurements of the temperature dependence of the longitudinal resistivity at
these filling factors. There are also some measurements about how the energy
gap itself changes with the magnitude of the magnetic field. The results are
mixed and it seems very difficult to conclude on a general physical pattern
from the multiple measurements. Obviously, more measurements are needed to
form a detailed microscopic picture. However, the availability of high quality
measurements in varied situations -- monolayer, bilayer, Corbino geometry etc.
-- in a wide range of sample quality and in both electron and hole branches,
is a rich resource to test and refine theories.

Many features of the QHE measured in graphene strengthen the case for the
gravitational paradigm of the QHE. Though graphene is fundamentally different
from the 2DEG system, the basic features of the IQHE and the FQHE in graphene
are consistent with the unified single particle scenario, in the relativistic
cosmic gravitational field induced by the dynamics of the charge carriers.

\section{Quantum-Thermal Effects in 2DEG and Graphene}

\begin{figure}
	\centering
	\includegraphics[width=0.5\linewidth]{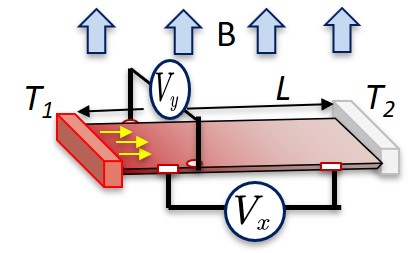}
	\caption{The Nernst effect and its measurement. The difference from the Hall effect measurement is that in this case a constant temperature gradient $\left( T_{1}-T_{2}\right) /L$ is maintained longitudinally, instead of using a constant current source. The charge transport is thermally driven. The transverse and the longitudinal voltages are measured under a magnetic field applied perpendicular to the 2D plane.}
	\label{fig:nerst}
\end{figure}

There is another relevant phenomenon at the interface of the 2D quantum
physics and thermodynamics, the Nernst effect, in which thermally driven
charged currents are Hall-deflected by the external magnetic field (figure \ref{fig:nerst}). When the
field is strong and the temperature is low, the quantized Landau levels
manifest in the periodic nature of the thermopower generated by the
temperature gradient. The phenomenon is well-understood \cite{Girvin-Nernst},
but the experiments are only a few. In the Corbino geometry, the azimuthal
Nernst thermocurrent is predicted to be oscillatory \cite{Alt-PNAS}. Graphene
is especially important for these studies because the quantum Hall effect
persists to relative high temperatures, which is essential to make good
measurements of the quantized thermopower \cite{Zuev-graph,Wei-graph}.
\emph{These measurements are important for testing the core theories, being
able to determine whether the magnetic field felt by the charge carriers is
the applied field or the one diminished by the hypothetical flux attachments
of the Composite Fermions}, for example. This needs better measurements at
lower temperatures and stronger magnetic fields, for the filling factors
$\nu<1$, to be conclusive. However, as already indicated by the standard
measurements of the FQHE in both 2DEG and graphene, I have no doubt that all
measurements related to the LL quantization will show results that
are consistent with the gravitational paradigm of the QHE; thermopower
measurements at the fractional filling factors, when available, will be no exception.

\section{Charge and Thermal Transport by the Edge States}

\begin{figure}
	\centering
	\includegraphics[width=0.9\linewidth]{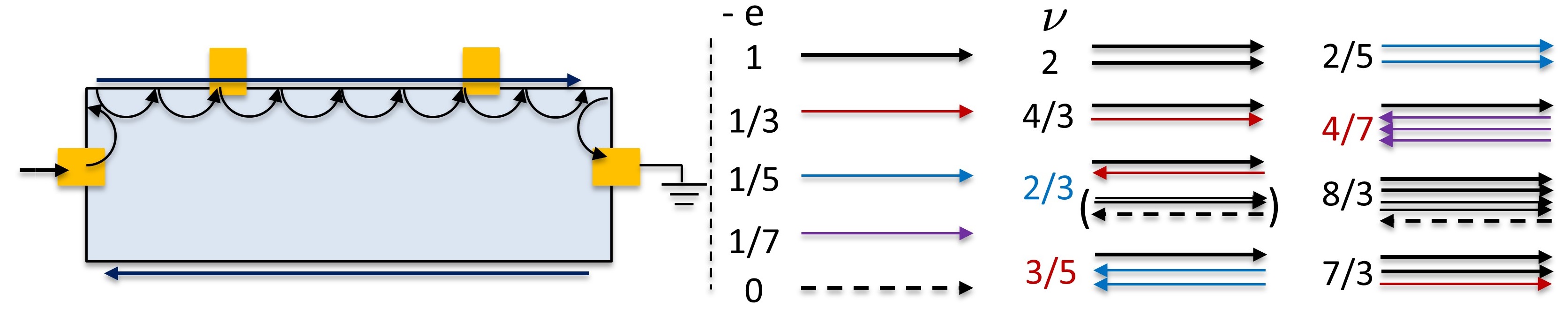}
	\caption{Left: A schematic depiction of edge state transport by the chirality
		preserving motion of the electrons in the large magnetic field, applied
		perpendicular to the 2DEG. The cyclotron motion at the edge is exaggerated in
		scale. The transport is robust and dissipationless because the general motion
		is unaffected by any scatterers near the edge. The yellow patches are the edge
		contacts for measurements. Right: The edge mode scheme in the hierarchical
		theories, consisting of the fractionally charged modes and a neutral mode. The
		`equilibration' modification to $2/3$ mode is shown in the brackets, which is
		also seen in the $8/3$ state. The cases where the heat flow is backward ($3/5$
		and $4/7$) are indicated with a different text colour. The general theoretical
		scheme is based on what is called the `K-matrix' formalism.}
	\label{fig:edge}
\end{figure}

So far I have been focussing on the peculiarities of the quantization
phenomena pertaining to the bulk of the 2DEG. However, the complexity of the
quantum Hall effects becomes more challenging when we consider transport
phenomena associated with currents confined to the \emph{edges} of the sample.
It was already indicated that the dissipationless conductance and the
vanishing longitudinal bulk resistance can coexist because of the transport
along the edges by what are called the `edge states', which are possible
because of the modification of the Landau energy levels near the edges of the
sample \cite{Halperin-edge}.  The dissipationless electrical current in the
QHE state is facilitated by these edge states \cite{Kane-edge}. These are
quantum mechanical versions of the unidirectional paths formed by the fixed
chirality of the cyclotron motion at the edges, determined by the direction of
the magnetic field (figure \ref{fig:edge}). Hence, the global directions of
the transport at the two edges are anti-parallel. The conductance per channel
is $e^{2}/h$ (spin polarized) and each Landau level contributes one
conductance channel.

Therefore, the quantized edge conductance is%
\begin{equation}
G=\frac{n_{c}e^{2}}{h}=\frac{\nu e^{2}}{h}%
\end{equation}
The explanation of the quantized conductivity becomes complex for the FQHE
state, in the prevalent hierarchical theories. However, there are adequate
descriptions, albeit involving multiple hypotheses, in the scenario invoking
the fractionally charged carriers and excitations, as well as in the composite
Fermion picture. These schemes invoke multiple forward (downstream) moving and
backward (upstream) moving edge state modes to provide an encompassing
description. Then the total quantized conductance is $G=\left(  e^{2}%
/h\right)  \sum v_{i}$, where $\nu_{i}$ is the filling fraction weight of each
channel. For example, the $\nu=2/3$ FQHE state has one forward moving edge
state mode of charge weight $-1e$ and one backward moving mode with weight
$e/3$ (in the same edge). This combination, instead of 2 forward channels of
weight $-1/3$, is forced by the hierarchical scheme of the theories. A state
like $\nu=4/7$ is more involved, with one forward mode with charge weight $-e$
and three backward modes of weight $e/7$ each. Such elaborate schemes can be
expected to explain the quantized fractional conductance at the edge. However,
the prediction for the edge state charge transport in the $\nu=2/3$ FQHE state
, $G_{2/3}=\left(  1+1/3\right)  e^{2}/h=4e^{2}/3h$, is never observed (the
$+1/3$ backward channel is equivalent to a forward $-1/3$ transport). Contrary
to this prediction, the observed conductance is $2e^{2}/3h$. This has
necessitated the introduction of a new hypothesis that there is inter-channel
tunneling interaction that makes the effective charge weightage $2e/3$ for the
forward mode. The backward mode is then forced to be a `neutral' (chargeless)
mode. Such a conversion from the bare modes to the mixed modes is supposed
to need a minimum equilibration distance through the sample, rendering the
detailed scenario complex. At the same time, the situation is very interesting
for experimental investigations because the edge modes can also transport
energy, in the form of `heat', if a temperature gradient exists along the
sample. Since it is the same charge carriers that also transport the energy,
the demand for consistency and coherence on the prospective theories become
stringent. Further, the scheme should be universal enough to be applicable to
QHE in both the 2DEG systems and graphene. None of the current theoretical
schemes is able to coherently cover the conflicting indications that emerge
from the limited experimental studies because of the need to consider thermal
transport as well, in additional to the electrical transport.

The experiments that study the thermal transport have reached the
sophistication and accuracy that challenge the theoretical description in any
simple scheme of edge state transport. Similar to the quantized electrical
transport for a ballistic channel, the thermal conductivity is known to be
quantized in units of $\pi^{2}k_{B}^{2}/3h$ \cite{Kane-Fisher}. In the early
predictions, the total conductance of thermal transport was
\begin{equation}
G_{T}=\left(  n_{+}-n_{-}\right)  \frac{\pi^{2}k_{B}^{2}}{3h}T\equiv
n_{c}\kappa T
\end{equation}
where $n_{+}$ and $n_{-}$ denote the number of forward and backward channels.
For the IQHE, $n_{-}=0$ and $n_{+}=\nu$, which agrees well with the numerous
measurements in 2DEG and graphene. The experiments in the FQHE states give
mixed results; some experiments are consistent with the hierarchical edge
state scenario of the current theories, whereas some others strongly disagree
with these theoretical schemes. While the $\nu=1/3$ state conducts energy is a
single channel ($n_{c}=1$), one expects no nett heat transport in the
$\nu=2/3$ FQHE state, in which the edge states are two counter propagating
modes. For a state like $\nu=4/7$, the number of backward modes (three) are
larger than the number of forward modes (one). Then one should expect the heat
flow at the edge in the backward direction, which is a surprising conclusion.

The gravitational paradigm and the unified theory of the quantum Hall effects
do not have the luxury of the hierarchical arrangements to adapt to each
situation of QHE. Each energy level that is fully filled to the degeneracy,
irrespective of whether it is a distinct Landau level or a level that is
determined by the smaller gap Coulomb interaction, can provide one edge
channel for transport. There are no fractionally charged quasiparticles or
composite particles in this theory. The transport is by single electrons in
1D, with separation determined by the density and the Coulomb interaction.
But, what is important is to realize that the transport coefficients are the
quantum expectation values (average); therefore, conduction in a level filled
to the modified degeneracy, because of the induced gravitational effect, is
transporting charge only at the factual number density and filling factor
$\nu$. In other words, the \emph{total charge weightage} remains exactly $\nu
$, even though the degeneracy is fully filled ($\nu^{\ast}$ is an integer, but
$\nu$ is fraction). The bulk conductivity and the edge conductivity match.
Therefore, the quantization of the electrical conductance at the edge is
intact as
\begin{equation}
G=\frac{\nu e^{2}}{h}%
\end{equation}

\begin{figure}
	\centering
	\includegraphics[width=0.7\linewidth]{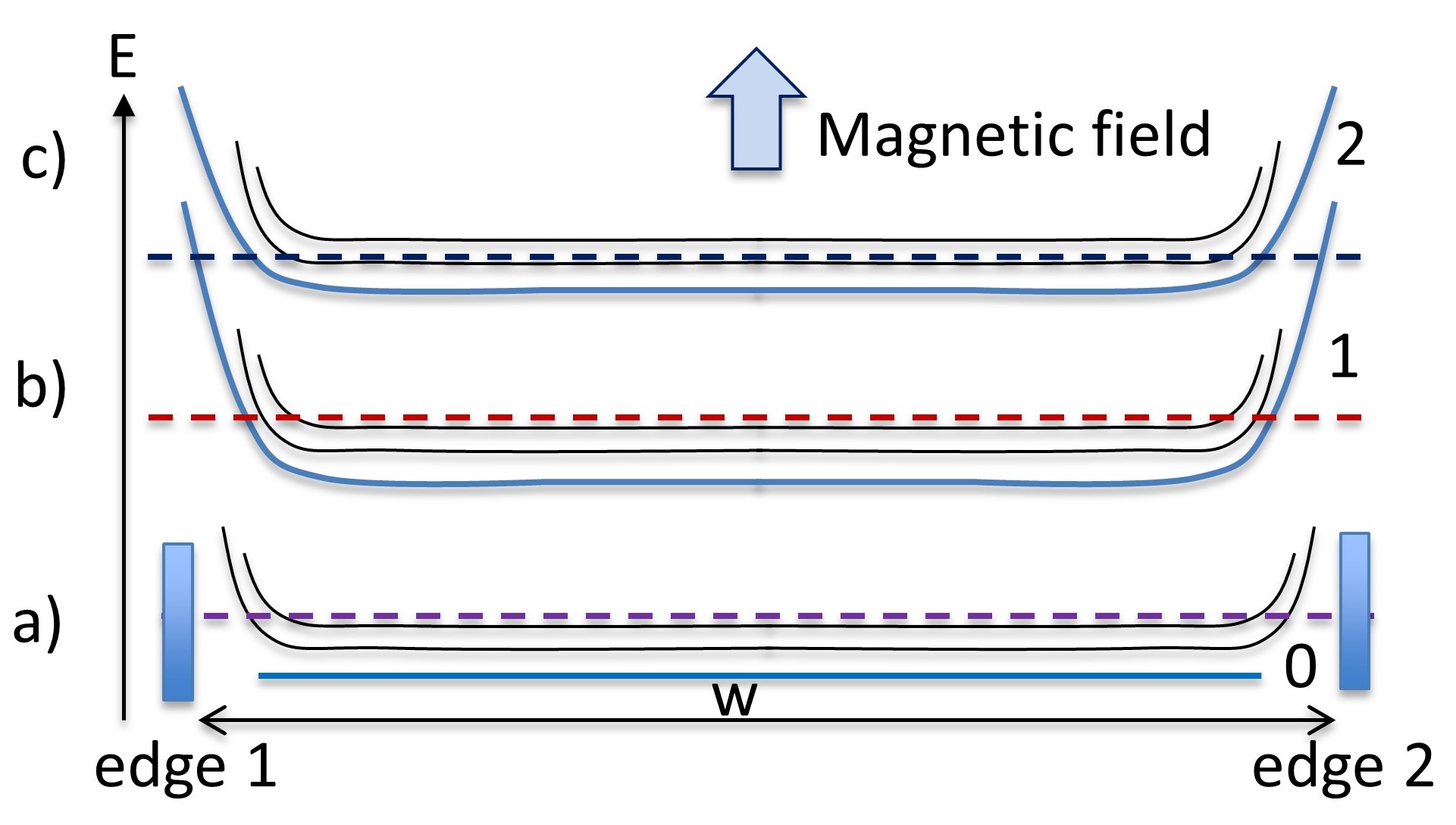}
	\caption{A schematic illustration and examples of the modification of the energy levels at the two Hall edges. Both the Landau levels as well as the Coulomb split sublevels are modified, but the exact nature of this modification is expected to vary for different levels. The edge channel of a level fully filled to its degeneracy (determined by both the magnetic field and induced cosmic gravitational field) contributes to conduction at its bare filling factor $\nu$. a) For the lowest level (0) and the filling factor 2/3, the effective degeneracy of each sublevel is 1/3 and there are 2 edge channels with the charge weightage 1/3 each. b) If the Fermi level is in the first LL, with filling factor 5/3, we can expect 3 edge channels. c) The situation is the same for the filling factor 7/3, but two of the 3 channels are provided by the LL 1 and 2, and the third (with charge weight 1/3) by a Coulomb sublevel. It is possible that some sublevels are not modified enough to contribute an edge channel.}
	\label{fig:edge-bend}
\end{figure}
We now analyse the thermal transport in the gravitational paradigm. The
conductance is by the real particles (electrons) that physically move along
the edge from one end of the sample to the other, constrained to be in their
chirality preserving motion in the large magnetic field. There is only one
chirality for a given configuration of the magnetic field and the factual
charge carriers. Therefore, there are no backward channels or neutral modes.
We expect each edge channel to contribute at most one unit of conductance,
depending on the extent of the modification of the levels near the edge. Thus,
all QHE state of $\nu<1$ may contribute one channel for each filled
Coulomb-split level. For example, for $\nu=1/3$, the degeneracy is exactly
filled with the effective filling factor $1$; then there is one conduction
channel, provide there is an edge state formed by the modification of the
potential at the edges. The available transport channels add for the thermal
conductance, in the absence of any other physical phenomena like the mixing
and the equilibration of the modes. Then, the thermal conductance in the QHE
situation is expected to be
\begin{equation}
G_{T}\leq\frac{\pi^{2}k_{B}^{2}}{3h}T\sum n_{i}\equiv\kappa Tn_{c}%
\end{equation}
where $n_{c}$ is the total number of edge state channels active in the
particular QHE state.

Before examining the experimental results, let is see what the gravitational
paradigm expects for the results of the measurements. The $\nu=p/q$ has at
most $p$ active forward edge modes, in which the conduction is by electrons.
Therefore $0\leq G_{T}\leq p\kappa T$. For $\nu>1$, a state $\nu=n+p/q$ can
have at most $n+p$ edge states, $n$ from the integer filling and $p$ from the
gravitationally completed Coulomb-split states (figure \ref{fig:edge-bend}). Therefore, the thermal
conductance is $n\kappa T\leq G_{T}\leq\left(  n+p\right)  \kappa T$. For
clean systems with sharp edges one then expects $G_{T}\simeq\left(
n+p\right)  \kappa T$.

The measurements are done in specially constructed mesoscopic devices with
access to the lithographically patterned edges through multiple contacts. The
exquisitely fabricated 2D devices were pioneered in an experiment by Jezouin
\textit{et al}. \cite{Jezouin2013}. Usually, a single mesoscopic resistance
heater sources the heat energy that can flow along the edges. There are point
contacts for the measurement of the charge and thermal conduction through
single channels. The thermal conduction is measured from the thermally induced
electronic noise in a resonant tuned circuit. Gate contacts to which
controlled bias voltages can be applied enable the selective conduction paths;
the information on how many channels are open at a particular bias voltage can
be obtained from the quantization of charge conductance in units of $e^{2}/h$.

There are only a few FQHE edge state thermometry experiments to date. The
recent experimental results on monolayer graphene \cite{AnindyaDas2021} are in
very good agreement with this prediction of the gravitational paradigm,
$G_{T}\simeq\left(  n+p\right)  \kappa$, and they do not support the
hierarchical theories unless additional hypotheses are introduced. The
conductivity measured in the states $\nu=5/3$, $\nu=7/3$, and $\nu=8/3$ gave
the results $G_{T}/\kappa T=3.03$, $2.96$ and $4.03$. Another set of
measurements in a different device for the states $\nu=4/3$, $\nu=7/3$, and
$\nu=8/3$ gave the results $G_{T}/\kappa T=1.96$, $3.01$ and $3.94$, in very
good agreement with the expectations.

However, the results of the measurements done in the 2DEG systems are mixed.
While all the results, for the charge transport as well as for the energy
transport, are consistent with the gravitational paradigm, they are also
partially consistent with the hierarchical theories. For example, in the
measurement reported in the reference \cite{Mitali2017}, $G_{T}/\kappa
T\simeq0.33$, $1.04$, and $2.04$ for the states $\nu=2/3$, $\nu=3/5$, and
$\nu=4/7$. In separate measurements, the same group reported $G_{T}/\kappa
T\simeq3$ for the state $\nu=7/3$, but for the state $\nu=8/3$, the result is
$G_{T}/\kappa T\simeq2$ \cite{Mitali2018}. It is still consistent with the
gravitational paradigm, in which $G_{T}/\kappa T=2$ and not $4$, if only the
Landau levels contribute to the edges transport, in any particular FQHE state.
In fact, then the $\nu=2/3$ state also does not contribute significantly to
the edge transport and we should get $G_{T}/\kappa T\simeq0$, as indeed seen
in the same device. Interestingly, for the state $\nu=5/2$, they measure
$G_{T}/\kappa T$ in the range $2.45$ to $2.75$, fully consistent with the
spatially mixed $\nu=\left(  2,3\right)  $ state advocated earlier in the
gravitational paradigm.

In summary, the new results in the graphene devices are in excellent agreement
with the gravitational paradigm, and explicitly contradict the minimal
schemes in the hierarchical theories. The measurements in the 2DEG systems
give mixed results that consistent with several schemes. However, the full
picture will take some time to emerge in a coherent manner. Certainly there is
the need to make more measurements, especially on the sign of the heat transport.

\section{The Wavefunctions in the QHE}

The single particle quantum states of the lowest Landau level are written as
\begin{equation}
\psi_{s}=z^{m}\exp\left(  -\left\vert z\right\vert ^{2}/4l_{0}^{2}\right)
\end{equation}
where $z=x+iy$ is called the coordinate of the electron. $m$ is the eigenvalue
of the angular momentum. The multi-particle `Laughlin wavefunction'
\cite{Laughlin-1983} built out of such single particle wavefunctions is
\begin{equation}
\Psi_{L}=\prod\limits_{i<j}\left(  z_{j}-z_{i}\right)  ^{m}\exp\left(
-\sum\limits_{i}\left(  \left\vert z_{i}\right\vert ^{2}/4l_{0}^{2}\right)
\right)
\end{equation}
with $m$ odd, which represents the filling factor $\nu=1/m$. The only
relevant fact I want to mention here is that this is in complete conformity
with the `degeneracy view' based on the dynamical action $S$, pursued in
this paper. The factor $z^{m}$ in the single particle wavefunction contains
the quantum phase factor, $z^{m}=r^{m}\exp\left(imS/\hbar\right)$. The
factor $S/h$ determines the quantum degeneracy, which in the case of the QHE
(or any noninertial dynamics), is determined by the applied fields and the
dynamics-induced cosmic gravitomagnetic field. Just as the action related to
the interaction of the charge with the magnetic potential determines the
quantum degeneracy, the action corresponding to the interaction of the mass
with the velocity dependent relativistic gravitational potential modifies the
degeneracy. In the multiparticle quantum mechanical context, the expectation
value of the degeneracy then becomes dependent on the actual particle number
density. For the quantum cyclotron dynamics, the action is contributed in the
ratio $1$ to $-2\nu$ by the magnetic field and the relativistic gravitational
field, revealing the true nature of the Laughlin wavefunction with $m=3$.
Thus, the factor $m$ here is not related to a fractional charge $1/m$ or to a
diminished effective magnetic field.

\section{Concluding Remarks}

I have presented a unified theoretical description of the integer and fraction
quantum Hall effects under a single encompassing paradigm of the factual
gravitational interaction of the electrons with the entire matter of this vast
Universe. Most strikingly, the dynamically induced cosmic gravitational potential modifies the quantum action and phase of the electrons
in their cyclotron orbitals, drastically modifying the degeneracy of the
Landau energy levels. The relativistic gravitational potential induced by the
motion of the electron is similar in nature to the magnetic vector
potential, but its coupling is to the mass of the particle. This interaction
modifies the quantum degeneracy in the magnetic field, $g_{m}=eB/h$,
to $g=g_{m}-2\rho$, \emph{making the degeneracy of the
levels dependent on both the magnetic field and the number density of the
charged particles}. This crucial result 
is the key to the seamless and unified single particle description of the
quantum Hall effects, without any quasiparticles or extraneous postulates. The
quantum degeneracy of the Landau levels determined by the two fundamental
interactions, $g=g_{m}(1-2\nu)$ where $\nu=\rho/g_{m}$ is the filling factor,
accounts for all the integer and prominent fractional Hall plateaux for
$\nu\geq1/3$. Also, obviously, it accounts for the absence of the QHE at the
even fraction $1/2$. The relativistically induced
gravitational field modifies only the quantum degeneracy, and the phase
factors in the wavefunctions, and not the nature of the charge carriers or the magnetic field experienced by them. The entire description is in terms of independent electrons, with
their mean Coulomb interaction at the factual charge density in the sample.

What is very important to realize is that irrespective of whatever
multi-particle electromagnetic effects are present in the condensed matter
situation, the cosmic gravitational effects discussed in this paper are
unavoidably present, because the matter in the Universe at the critical
cosmological density is observationally verified and it has the corresponding
gravity. 

I have commented on plausible reasons that might account for the
fractions for $\nu<1/3$, within the same paradigm, but this needs 
additional postulates concerning the dynamics. Also, I have conjectured that the observed $5/2$ even FQHE is in fact
a mixed phase of $\nu=2$ and $\nu=3$ FQHE (along with $\nu=7/3$ and $\nu=8/3$).

I have compared the QHE in 2DEG and monolayer graphene, two widely different physical environments, to emphasize the universal aspects of the QHE. In particular, the importance of the thermal and charge transport studies in different geometries was highlighted. I analysed the experimental studies on the edge state transport in the context of the cosmic gravitational paradigm. It was found that the results obtained with monolayer graphene samples are in very good agreement with the expectations. 

What is achieved in this work is both the unification of the observed quantum
Hall phenomena in 2-d materials under a single paradigm, crossing
expectations, and the unravelling of the precise physical mechanism, in a
transparent way that can be verified. In an experimentally intense field, with
many remarkable results, there is no last word yet regarding the details. But,
the essential reason for the FQHE, and its quantitative determining factors,
is now clear. \emph{The FQHE and IQHE are the same phenomenon, in which
electromagnetism and gravity have equal but different roles}. The cosmic matter
and its gravity are bare facts that we do not assimilate because of the accident
of the relative history of physics and cosmology; the cart went quite ahead
before the horse was born. But cosmic gravity has well-defined consequences to
noninertial dynamics through the relativistic gravitational potentials. Though
a complete understanding of such a complex phenomenon with many details, some
yet to discover, is beyond any theory of the essentials, what is guaranteed is
that no empirical fact will go against the unified theory of the quantum Hall
effects within the cosmic gravitational paradigm.

While being struck by the beauty of the night sky with innumerable stars,
let us also realize that the immense amount of the observationally confirmed
cosmic matter has an equally immense gravity, which dominates the dynamics of
every particle. The Universe is not a passive backdrop to laboratory physics,
but the determining factor in many situations as the discussion in this paper
shows. The importance of cosmic gravity can be verified even by the simplest
of calculations of the gravitational influence on moving clocks and quantum
systems, due to the matter in the nearby galactic cluster. Going further, when
we consider the whole matter-filled Universe, cosmic gravity becomes the
determining factor of dynamics itself \cite{Unni-Cosrel,Unni-PhysNews,GravityTime}. That cosmic gravity
is the key for understanding one of the most intriguing discoveries of recent
times, the fractional quantum Hall effect, might be
unbelievably surprising to most. But, it is indeed one of the many
enlightening examples of the encompassing
role of cosmic gravity in relativistic dynamics and quantum mechanics.

\begin{acknowledgments}
I am thankful to Jainendra Jain for patiently explaining various aspects of
the Composite Fermion theory of the fractional Quantum Hall effect. Martine
Armand helped in improving the structure and flow of this article.
\end{acknowledgments}

\end{document}